\documentclass[12pt,preprint]{aastex}
\usepackage{emulateapj5,apjfonts}
\usepackage{graphics}
\usepackage{amssymb}
\usepackage{bbm}
\usepackage{amsmath,textcomp,array}
\usepackage{onecolfloat}
\usepackage{bm}

\lefthead{Heitmann et al.}
\righthead{The Coyote Universe Extended}

\begin{document}


\def\head{

\title{The Coyote Universe Extended: \linebreak Precision Emulation of
  the Matter 
  Power Spectrum}
\author{Katrin~Heitmann\altaffilmark{1,2,3,4}, Earl
  Lawrence\altaffilmark{5}, 
Juliana Kwan\altaffilmark{1}, Salman~Habib\altaffilmark{1,2,3,4},  
\linebreak and David Higdon\altaffilmark{5}}

\affil{$^1$ High Energy Physics Division, Argonne National Laboratory,
 Lemont, IL 60439}
\affil{$^2$ Mathematics and Computer Science Division, Argonne
  National Laboratory, 
 Lemont, IL 60439}
\affil{$^3$ Kavli Institute for Cosmological Physics, The University
  of Chicago, Chicago, IL 60637 } 
\affil{$^4$ Computation Institute, The University of Chicago, Chicago,
  IL 60637} 
\affil{$^5$ CCS-6, CCS Division, Los Alamos National Laboratory, Los
  Alamos, NM 87545} 

\date{today}

\begin{abstract}

  Modern sky surveys are returning precision measurements of
  cosmological statistics such as weak lensing shear correlations, the
  distribution of galaxies, and cluster abundance. To fully exploit
  these observations, theorists must provide predictions that are at
  least as accurate as the measurements, as well as robust estimates
  of systematic errors that are inherent to the modeling process. In
  the nonlinear regime of structure formation, this challenge can only
  be overcome by developing a large-scale, multi-physics simulation
  capability covering a range of cosmological models and astrophysical
  processes. As a first step to achieving this goal, we have recently
  developed a prediction scheme for the matter power spectrum (a
  so-called {\it emulator}), accurate at the 1\% level out to
  $k\sim1~$Mpc$^{-1}$ and $z=1$ for $w$CDM cosmologies based on a set
  of high-accuracy N-body simulations. It is highly desirable to
  increase the range in both redshift and wavenumber and to extend the
  reach in cosmological parameter space. To make progress in this
  direction, while minimizing computational cost, we present a
  strategy that maximally re-uses the original simulations. We
  demonstrate improvement over the original spatial dynamic range by
  an order of magnitude, reaching $k\sim10~h$Mpc$^{-1}$, a four-fold
  increase in redshift coverage, to $z=4$, and now include the Hubble
  parameter as a new independent variable. To further the range in $k$
  and $z$, a new set of nested simulations run at modest cost is added
  to the original set. The extension in $h$ is performed by including
  perturbation theory results within a multi-scale procedure for
  building the emulator. This economical methodology still gives
  excellent error control, $\sim 5\%$ near the edges of the domain of
  applicability of the emulator. A public domain code for the new
  emulator is released as part of the work presented in this paper.

\end{abstract}

\keywords{methods: statistical ---
          cosmology: large-scale structure of the universe}}

\twocolumn[\head]

\section{Introduction}
\subsection{Background}

The discovery of the accelerated expansion of the Universe by
 \cite{riess} and  \cite{perlmutter}, now confirmed by multiple
observations, has had a tremendous impact on the field of
observational and theoretical cosmology. Due to our near-complete
ignorance regarding the origin of cosmic acceleration, the initial
task is to better characterize the nature of the underlying cause,
first by observational means. The primary focus in this endeavor is to
decide whether the acceleration is caused by some form of dark energy,
a cosmological constant being the simplest realization of this idea,
or by some modification of general relativity on large cosmological
scales.

Unfortunately, carrying out controlled experiments in cosmology is
impossible -- therefore, information relevant to the above task must
be obtained by combining inferences from different kinds of
observations. These observations include the measurement of the
temperature anisotropy of the cosmic microwave background (CMB), the
luminosity distance relation from supernova observations, and large
scale structure probes, such as baryon acoustic oscillations (BAO),
weak lensing, clustering of galaxies, and the abundance of galaxy
clusters. Despite the impressive power of the observations, all
cosmological probes are plagued with a large number of sources of
systematic error. These can be broken up into two kinds, (i) those
related to the observational techniques -- shape measurements in weak
lensing and the determination of cluster masses being two prominent
examples, and (ii) those related to the uncertainties in the
theoretical predictions and the (related) errors in solving
cosmological statistical inverse problems. An important member of this
latter class is the prediction accuracy of the matter power spectrum,
the central theme of this paper.

Because robust results in cosmology must rely on combining a number of
observational probes, each possessing their individual strengths and
weaknesses (not all of which can be fully predicted in advance), the
current strategy is to proceed with several survey missions. In the
spectroscopic realm, ongoing or recently completed surveys focusing on
the measurement of BAO and galaxy clustering include the Sloan Digital
Sky Survey (SDSS) \citep{zehavi}, the WiggleZ survey \citep{wigglez}
and the Baryon Oscillation Spectroscopic Survey~(BOSS) \citep{boss}
component of SDSS-III; planned next-generation missions include the
Mid-Scale Dark Energy Spectroscopic Instrument (MS-DESI) and
SuMIRe. Imaging surveys include SDSS, the Canada-France-Hawaii
Telescope Legacy Survey (CFHTLS), the Kilo Degree Survey (KiDS)~(\citealt{kids}), 
the recently started Dark Energy
Survey (DES), and in the future, Subaru Measurement of Images and
Redshifts (SuMIRe) \citep{sumire}, the Large Synoptic Survey Telescope
(LSST) \citep{lsst}, and Euclid \citep{euclid}. The most recent
constraints from weak lensing have been obtained by the CFHTLenS team
\citep{cfhtlens}, analysis of SDSS data (\citealt{huff},
\citealt{lin}), and the Deep Lens Survey~(DLS) (\citealt{dls}). In the x-ray the next planned mission is eROSITA
\citep{erosita}, which will provide a significant update to results
from prior surveys, such as the 400d cluster survey
\citep{vikhlinin}. In the CMB, high-resolution measurements by the
Atacama Cosmology telescope (ACT) \citep{act} and the South Pole
Telescope (SPT) \citep{spt} have confirmed the sensitivity of the CMB
to the large scale distribution of matter (CMB lensing).
Significantly, first results from the Planck mission have been
recently released \citep{planck}.  For a comprehensive review of the
current state-of-the-art (pre-Planck) we refer the reader to
\cite{weinberg}.

With the advent of ambitious large-scale observational programs,
emphasis on statistical errors will no longer be the sole driving
force. The understanding, control, and reduction of systematic errors
is the next major challenge -- there are several cases where
measurements are already systematics-limited. Additionally, in the
nonlinear regime of structure formation, state of the art measurements
are pushing up against the limits of available prediction accuracy
from theoretical computations and associated modeling. Major
simulation campaigns, in concert with observational inputs, will be
needed to control the error budgets. The simulation-based theoretical
modeling process -- consisting of a combination of first-principles
calculations and parameterized models -- will have to account
accurately, and robustly, for the nonlinear evolution of structure
formation, gas physics, and complex feedback effects.

Two examples will suffice to provide illustrations of the sort of
difficulties that have to be faced. The first is cluster cosmology --
it was recently shown by \cite{cunha10} and  \cite{wu} that the halo
mass function has to be predicted at the 1\% level of accuracy for a
DES-like survey in order to avoid errors on dark energy parameters. At
the same time, an error in the prediction of halo masses at the 2\%
level will translate into mass function uncertainties in the cluster
mass range of 5-10\% (Cf.  \citealt{bhattacharya11}). The effect is
particularly strong because of the exponential fall-off of the mass
function at high masses. The second example relates to prediction
error requirements for weak lensing shear. These are also severe --
for the matter power spectrum, predictions at the percent level
accuracy at $k\sim 10~h^{-1}$Mpc are needed to extract all available
information on dark energy \citep{huttak}. A more recent study by
 \cite{hearin} concludes that the accuracy requirements are even more
stringent -- for imaging surveys such as LSST and Euclid, the power
spectrum needs to be predicted at 0.5\% accuracy out to $k\sim
5~h^{-1}$Mpc.

\subsection{Fitting Functions vs. Emulators}

Currently, it is common practice to use fitting functions motivated by
theoretical heuristics and matched to simulations to provide the
required modeling input. For instance, analyses of weak lensing
measurements employ fitting functions for the matter power
spectrum. \cite{semboloni} used the Peacock and Dodds fitting function
\citep{PD96} as well as Halofit \citep{Smith03} for their analysis of
the CFHTLS cosmic shear measurements. \cite{lin} used Halofit for
their recent SDSS analysis, as do \cite{cfhtlens} in the most recent
CFHTLens analysis (they point out that an emulator would be more
accurate but one was not available over the required redshift range)
and \cite{dls} for the DLS analysis.
Over a limited $k$ range, the fitting functions are accurate at the
$\sim 5-10\%$ level for $\Lambda$CDM models (see,
e.g., \citealt{coyote1}). The latest improvement of Halofit is provided
by \citealt{takahashi}; a comparison of our results with theirs is
provided in Section~\ref{halofit}.

For next-generation applications, the fitting function approach
suffers from two deficiencies, first, error-control is non-uniform
over cosmological model parameter space, and, second, the fitting
process degrades the actual accuracy of the simulations used to build
the parameterized fit. Moreover, because the fitting forms become
essentially arbitrary as the accuracy requirements become more
stringent, this strategy is difficult -- if not impossible -- to
implement systematically across a large number of freely floating
cosmological and modeling parameters.

Cluster mass functions provide another example of these difficulties
-- obtaining dark energy constraints from the abundance of clusters of
galaxies as carried out in, e.g., \cite{vikhlinin}, employs fitting forms for the mass
function; to extend these fits across cosmological models, the
assumption of universality  \citep{jenkins01} is required. However, as
shown in, e.g., \cite{bhattacharya11}, universality for $w$CDM models
only holds at the $\sim10$\% level of accuracy. It appears quite
unlikely that a simple mass function fit -- valid for a large class of
dark energy models, covering a wide redshift range, and satisfying
percent level accuracy requirements -- will ever be attainable.

We have recently developed the ``Cosmic Calibration Framework'' (CCF)
to provide accurate prediction schemes for cosmological observables
(\citealt{HHHN}, \citealt{HHHNW}) seeking to avoid the shortcomings of
the fitting function approach mentioned above. The CCF contains
error-controlled direct numerical oracles for the predicted
quantities, as opposed to (potentially uncontrolled) analytic fitting
forms. It aims to make tools available that provide
essentially instantaneous predictions of large scale structure
observables, e.g., the nonlinear power spectrum, mass functions for
different halo definitions, or the halo concentration-mass relation.

At the heart of the CCF lies a sophisticated sampling scheme for
optimally placing a given finite number of cosmological models in
parameter space. Simulations are carried out at these parameter values
(we use orthogonal-array Latin hybercubes as well as symmetric Latin
hybercube designs). The next step is an efficient representation that
translates the measurements from the simulations into functions that
are conveniently interpolated (we use a principal component basis to
provide a reduced data representation), and finally an accurate
interpolation scheme over the basis functions (our choice here is
Gaussian process modeling). For an introduction to the general
framework, see, e.g., \citealt{santner03}.

The CCF was first described in \cite{HHHN}, with details and examples
given in \cite{HHHNW}. In a series of three papers (Coyote Universe
I-III), we described the development of a precision emulator for the
matter power spectrum over a five-dimensional parameter space
($\theta=\{\omega_b, \omega_m, n_s, w, \sigma_8\}$). This emulator
provides predictions for the power spectrum for $w$CDM cosmologies out
to $k\sim 1$~Mpc$^{-1}$ at the 1\% accuracy level for a redshift range
of $0\le z\le 1$, over a relatively wide range of cosmological
parameters (see Section~\ref{sec:models}). Since then, similar
approaches have been followed in \cite{schneider} and \cite{feldman}
with some differences in the sampling and interpolation schemes used
to build power spectrum emulators.

The CCF framework was extended in \cite{SKHHHN} to derive an
approximate statistical model for the sample variance distribution of
the nonlinear matter power spectrum. \cite{eifler} used the emulator
to generate a weak-lensing prediction code to calculate various
second-order cosmic shear statistics, e.g., shear power spectrum,
shear-shear correlation function, ring statistics and Complete
Orthogonal Set of EB-mode Integrals (COSEBIs). The emulator we presented
in \cite{coyote3} was used by \cite{huff} in their analysis of SDSS
weak lensing (where it was combined with Halofit to obtain the power
spectrum at larger $k$ values). More CCF-based emulators are under
development; a recent example is an emulator for the halo
concentration-mass relation \citep{kwan12}.

Observational requirements, such as those for DES weak lensing, and
the work by  \cite{eifler} and  \cite{huff} have prompted us to extend
the  \cite{coyote3} power spectrum emulator in three directions: (i)
making the Hubble parameter $h$ freely choosable (in the original
emulator the value for $h$ in each model is fixed by the CMB distance
to last scattering constraint); (ii) extending the $k$ range out to
$k=8.6~$Mpc$^{-1}$; (iii) extending the $z$ range out to $z=4$. As
described in Section~\ref{sec:models} below, these extensions are
nontrivial -- a brute force approach would require simulations of size
$L=1.5-2$~Gpc with at least $N_p^3=10,000^3$ particles to fulfill the
requirements for 1\% accuracy, as derived in  \cite{coyote1}. The most
stringent requirement is due to the fact that the power spectrum
should not be measured further out than half of the particle Nyquist
frequency, $k_{\rm Ny}=\pi/\Delta_p$ with $\Delta_p$ being the initial
particle separation $L/N_p$, and at the same time the linear modes
should be well resolved in the simulation volume. In addition, at high
redshift and large $k$, particle shot noise becomes a significant
issue (see the discussion in Section~\ref{sec:models}).

Running a large number of the brute force simulations described above
is currently impractical. In order to extend our prediction scheme we
therefore choose another route, by using a set of nested simulations
of various volumes, and matching the results together at different
values of $k$ and $z$. The box lengths are chosen to be $L=1300$~Mpc
(these are the simulations from the original Coyote Universe project
as described in \citealt{coyote3}) and $L=365$~Mpc below
$z\sim0.7$. For higher redshifts and smaller scales we use simulations
of size $L=180$~Mpc, and for redshifts $z>2$, $L=90$~Mpc.\footnote{The
  intermediate box sizes of $L=365$~Mpc and $L=180$~Mpc lead to good
  mass resolution and have also been used to determine the
  concentration-mass relation for $w$CDM cosmologies by
  \cite{kwan12}.} The specific choices for the box size are explained
in more detail in Section~\ref{sec:models}. In short, these choices
ensure good overlap between the resulting power spectra at different
scales and redshifts and limit inaccuracies due to shot noise and
finite box size effects.  

A nested approach in the same spirirt, and with similar matching
scales in $k$ and redshift, has been successfully employed by
\cite{takahashi}. Based on their simulation results, these authors
have developed an improved version of Halofit for predicting the
nonlinear power spectrum. We compare their results to ours in detail
in Section~\ref{halofit}. The nested box approach has its
deficiencies, as discussed further in
Section~\ref{sec:models}. Nevertheless, carrying out a battery of
tests we can demonstrate that our predictions hold at the level of
better than 5\% error over the full $k$
range. Appendix~\ref{appendixa} shows one example of a large, high
resolution simulation that spans most of the $k$-range covered by the
new emulator and the comparative error is $\sim\pm2\%$.

\subsection{The Power Spectrum at Small Scales: Baryonic Effects}

At small length scales, complex baryonic processes become
important. Because these are difficult to model, a relatively large
uncertainty in the power spectrum exists over the upper $k$ range
treated in this paper. (Massive neutrinos also have a small effect,
see, e.g.,  \citealt{bird}.)  \cite{white04} estimated the effect of
baryonic cooling on the mass power spectrum using the halo model,
finding an increase in power at $k\sim 10~h$Mpc$^{-1}$ at the level of
a few percent at $z=0$. An analogous approach by \cite{zhanknox} to
study the effect of the intracluster medium (hot baryons) found a
suppression in the power spectrum at roughly similar levels.

Shortly after these papers appeared, simulations were carried out to
study these effects in more detail.  \cite{jing06} carried out two
simulations with the smoothed-particle hydrodynamics (SPH) code {\sc
  Gadget-2}, a non-radiative gas simulation and another including
radiative cooling and star formation, with a third gravity-only
simulation serving as a reference. At $k\simeq 10~h$Mpc$^{-1}$ they
found an effect on the total matter power spectrum at the few percent
level in the non-radiative gas simulation and at the 10\% level in the
gas simulation with radiative cooling and star formation at $z=0$ (the
effects are smaller at higher redshifts -- for the radiative
cooling/star formation simulation, they are at the 2-3\% level at
$z=1$). In both cases, the baryonic effects led to an enhancement in
the power spectrum. A similar campaign was later carried out by
\cite{rudd}: gravity-only, non-radiative gas dynamics, and a third
simulation including radiative heating and cooling of baryons, star
formation, and supernova feedback. This set of simulations used the
Adaptive Refinement Tree (ART) code, combining an N-body method and an
Eulerian hydrodynamics solver on an adaptive mesh. For the
non-radiative gas simulation they found a similar effect as
\cite{jing06} on the overall matter power spectrum, though effects on
the baryon and dark matter components separately were different. The
results from their third simulation showed a dramatic increase of the
matter power spectrum well beyond the effect that \cite{jing06}
observed (at the 50\% level at $k \simeq 5~h$Mpc$^{-1}$) -- but as
noted by the authors, this result was not meant to be definitive.

More recent studies have been carried out by  \cite{vanDaalen}, again
using {\sc Gadget-2} (the authors also provide a much more
comprehensive discussion of results from other groups than we have
space to do here).  The main difference in this study is the inclusion
of feedback from active galactic nuclei (AGN). They observe a strong
effect on the power spectrum in the opposite direction than found by
the previous studies: a decrease in the matter power spectrum at 1\%
at $k\simeq 0.3~h$Mpc$^{-1}$, at 10\% at $k\simeq 1~h$Mpc$^{-1}$, and
at 30\% at $k\simeq 30~h$Mpc$^{-1}$. As shown by these results,
obtaining accurate first principles predictions for the power spectrum
including baryonic effects will be very difficult. Multiple issues,
ranging from physical effects to simulation uncertainties, remain to
be sorted out.

A strategy to properly deal with these difficulties is still emerging,
but it will definitely require methods to incorporate information from
observations, baryonic simulations, and an accurate calibration of the
gravity-only power spectrum (which has no free modeling
parameters). The difficulty of accounting for all ``gastrophysics''
effects at the desired accuracies, almost certainly precludes using
hydro simulations in a fully predictive mode. One possible pathway is
to construct parameterized models for baryonic effects on top of
predictive N-body simulation results (see e.g., \citealt{semboloni11,
  zentner12} for some recent work) and combine these with (possibly
multiple) observations (`self-calibration'), to successfully extract
cosmological information. In this paper, our target is to establish
the bedrock on which all such analyses must rest -- an accurate
calibration of the gravity-only matter power spectrum.

The rest of the paper is organized as follows. After providing a brief
summary of our power spectrum estimation from the simulations
(Section~\ref{power}), we describe the cosmological model space
covered and the simulation suite used to build the emulator
(Section~\ref{sec:models}). In Section~\ref{sec:hubble} we outline our
strategy to include the Hubble parameter $h$ as a free parameter,
without adding more N-body simulations. Next we discuss the generation
of smoothed predictions for the power spectra for each model; this
process underlies the interpolation scheme for constructing the
emulator. In Section~\ref{sec:emu} we show some examples from the
working emulator, including a brief comparison with Halofit. Finally,
we end with a conclusion and outlook in Section~\ref{sec:conc}. 
Appendix B~\ref{appendixb} presents a short discussion of an improvement
to the general Gaussian process approach employed here compared to
that in the previous Coyote papers, including a minor numerical
correction of the earlier results.

\section{Power spectrum estimation} 
\label{power}

The key statistical observable in this paper is the density
fluctuation power spectrum $P(k)$, the Fourier transform of the
two-point density correlation function. We follow the same strategy
for extracting the power spectra from the simulations as
in \cite{coyote1} and give a summary here for completeness.

In dimensionless form, the power spectrum may be written as
\begin{equation}
  \Delta^2(k) \equiv \frac{k^3P(k)}{2\pi^2},
\label{delta}
\end{equation}
which is the contribution to the variance of the density
perturbations per $\ln k$.  

Because $N$-body simulations use particles, one does not directly
compute $P(k)$ or equivalently, $\Delta^2(k)$. Our procedure is to
first define a density field on a grid of size $N_g^3$ with a fine
enough resolution such that the grid filtering scale is much higher
than the $k$ scale of interest.  This particle deposition step is
carried out using Cloud-in-Cell (CIC) assignment. The application of a
discrete Fourier transform (FFT) then yields $\delta(\bf{k})$ from
which we can compute $P(\bf{k})=|\delta(\bf{k})|^2$, which in turn can
be binned in amplitudes to finally obtain $P(k)$. Since the CIC
assignment scheme is in effect a spatial filter, the smoothing can be
compensated by dividing $P(\bf{k})$ by $W^2(\bf{k})$, where
\begin{equation}
  W\left(\bf{k}\right) =
    j_0^2\left(\frac{k_xN_g}{2}\right)
    j_0^2\left(\frac{k_yN_g}{2}\right)
    j_0^2\left(\frac{k_yN_g}{2}\right).
\end{equation}
Typically the effect of this correction is only felt close to the
maximum (Nyquist) wavenumber for the corresponding choice of grid
size. One should also keep in mind that particle noise and aliasing
artifacts can arise due to the finite number of particles used in
$N$-body simulations and due to the finite grid size which is used for
the power spectrum estimation, as discussed further in
Section~\ref{sec:nested}. For an extensive suite of convergence tests
addressing these issues, see, e.g., \cite{coyote1}.

We average $P(\bf{k})$ in bins linearly spaced in $k$ of width $\Delta
k\simeq 0.001\,{\rm Mpc}^{-1}$, and report this average for each bin
containing at least one grid point.  We assign to each bin the $k$
associated with the unweighted average of the $k$'s for each grid
point in the bin.  Note that this procedure introduces a bias in
principle, since for nonlinear functions $\langle f(x)\rangle\ne
f(\langle x \rangle)$, but our bins are small enough to render this
bias negligible. Throughout the paper we will show results for both
$\Delta^2(k)$ and $P(k)$. The emulator itself provides results for
both definitions as well.

\section{Cosmological Models and Simulation Sets}
\label{sec:models}

The emulator is based on 37 cosmological models spanning the class of
$w$CDM cosmologies. We allow for variations of the following six
parameters:
\begin{equation}\label{eq:theta}
\theta=\{\omega_b, \omega_m, n_s, h, w, \sigma_8\}.
\end{equation}
The 37 models are chosen to lie within the ranges:
\begin{equation}
\begin{array}{c}
0.0215 < \omega_b < 0.0235, \\
  0.120 < \omega_m < 0.155, \\
  0.85  < n_s      < 1.05, \\
  0.55 < h <  0.85,\\
  -1.30 < w        <-0.70, \\
  0.616  < \sigma_8 < 0.9,
\end{array}
\label{priors}
\end{equation}
motivated by recent constraints from CMB
measurements by WMAP \citep{wmap7}. 

Since our work is mainly aimed at current and near-future weak lensing
measurements, we restrict our model space to $w$CDM cosmologies, not
considering dynamical dark energy models. We will address these models
in future work (for a more detailed discussion regarding the rationale
behind the cosmological model choices see \citealt{coyote2}).

In our original work we locked the value of the Hubble parameter $h$
to the best-fit value given a measurement of the distance to the
surface of large scattering for each model \citep{coyote3}. The values
for $h$ in the original runs then ranged from $0.55 < h < 0.85$. While
this range is larger than the currently available constraints on the
Hubble parameter from, e.g., \cite{riess_hubble} we decide to keep it
in our new runs in order to include the best fit models. In
Section~\ref{sec:hubble} we explain how we extend the parameter range
to include $h$ as a free variable in the new emulator without actually
running more N-body simulations. In addition to the 37 models, we ran
one $\Lambda$CDM model (M000 in Table~\ref{tab:basic}) which is not
used to build the emulator, but is instead used as a reference to test
the emulator accuracy. All 37+1 models are specified in detail in
Table~\ref{tab:basic}.

\begin{table*}
\begin{center} 
\caption{Parameters for the 37+1 models which define the sample
  space; $k_{\rm nl}$ is measured in Mpc$^{-1}$. See text for further
  details. \label{tab:basic}} 
\vspace{-0.3cm}
\begin{tabular}{ccccccc|ccccccc}
\# & $\omega_m$ & $\omega_b$ & $n_s$ & $-w$ & $\sigma_8$ & $h$ &
\# & $\omega_m$ & $\omega_b$ & $n_s$ & $-w$ & $\sigma_8$ & $h$ \\ \hline 
 M000 & 0.1296 & 0.0224 & 0.9700 & 1.000 & 0.8000 & 0.7200  &
M019 & 0.1279 & 0.0232 & 0.8629 & 1.184 & 0.6159 & 0.8120  \\ 
 M001 & 0.1539 & 0.0231 & 0.9468 & 0.816 & 0.8161 & 0.5977 & 
M020 & 0.1290 & 0.0220 & 1.0242 & 0.797 & 0.7972 & 0.6442  \\ 
 M002 & 0.1460 & 0.0227 & 0.8952 & 0.758 & 0.8548 & 0.5970 & 
M021 & 0.1335 & 0.0221 & 1.0371 & 1.165 & 0.6563 & 0.7601  \\ 
 M003 & 0.1324 & 0.0235 & 0.9984 & 0.874 & 0.8484 & 0.6763 & 
M022 & 0.1505 & 0.0225 & 1.0500 & 1.107 & 0.7678 & 0.6736  \\ 
 M004 & 0.1381 & 0.0227 & 0.9339 & 1.087 & 0.7000 & 0.7204 & 
M023 & 0.1211 & 0.0220 & 0.9016 & 1.261 & 0.6664 & 0.8694 \\ 
M005 & 0.1358 & 0.0216 & 0.9726 & 1.242 & 0.8226 & 0.7669 & 
M024 & 0.1302 & 0.0226 & 0.9532 & 1.300 & 0.6644 & 0.8380 \\ 
M006 & 0.1516 & 0.0229 & 0.9145 & 1.223 & 0.6705 & 0.7040 & 
M025 & 0.1494 & 0.0217 & 1.0113 & 0.719 & 0.7398 & 0.5724 \\ 
 M007 & 0.1268 & 0.0223 & 0.9210 & 0.700 & 0.7474 & 0.6189 &
M026 & 0.1347 & 0.0232 & 0.9081 & 0.952 & 0.7995 & 0.6931  \\ 
 M008 & 0.1448 & 0.0223 & 0.9855 & 1.203 & 0.8090 & 0.7218 & 
M027 & 0.1369 & 0.0224 & 0.8500 & 0.836 & 0.7111 & 0.6387  \\ 
 M009 & 0.1392 & 0.0234 & 0.9790 & 0.739 & 0.6692 & 0.6127 & 
M028 & 0.1527 & 0.0222 & 0.8694 & 0.932 & 0.8068 & 0.6189  \\ 
M010 & 0.1403 & 0.0218 & 0.8565 & 0.990 & 0.7556 & 0.6695 & 
M029 & 0.1256 & 0.0228 & 1.0435 & 0.913 & 0.7087 & 0.7067  \\ 
M011 & 0.1437 & 0.0234 & 0.8823 & 1.126 & 0.7276 & 0.7177 & 
M030 & 0.1234 & 0.0230 & 0.8758 & 0.777 & 0.6739 & 0.6626  \\ 
M012 & 0.1223 & 0.0225 & 1.0048 & 0.971 & 0.6271 & 0.7396 & 
M031 & 0.1550 & 0.0219 & 0.9919 & 1.068 & 0.7041 & 0.6394  \\ 
M013 & 0.1482 & 0.0221 & 0.9597 & 0.855 & 0.6508 & 0.6107 & 
M032 & 0.1200 & 0.0229 & 0.9661 & 1.048 & 0.7556 & 0.7901  \\
M014 & 0.1471 & 0.0233 & 1.0306 & 1.010 & 0.7075 & 0.6688 & 
M033 & 0.1399 & 0.0225 & 1.0407 & 1.147 & 0.8645 & 0.7286  \\ 
M015 & 0.1415 & 0.0230 & 1.0177 & 1.281 & 0.7692 & 0.7737 & 
M034 & 0.1497 & 0.0227 & 0.9239 & 1.000 & 0.8734 & 0.6510  \\  
M016 & 0.1245 & 0.0218 & 0.9403 & 1.145 & 0.7437 & 0.7929 & 
M035 & 0.1485 & 0.0221 & 0.9604 & 0.853 & 0.8822 & 0.6100  \\ 
M017 & 0.1426 & 0.0215 & 0.9274 & 0.893 & 0.6865 & 0.6305 & 
M036 & 0.1216 & 0.0233 & 0.9387 & 0.706 & 0.8911 & 0.6421  \\
M018 & 0.1313 & 0.0216 & 0.8887 & 1.029 & 0.6440 & 0.7136 & 
M037 & 0.1495 & 0.0228 & 1.0233 & 1.294 & 0.9000 & 0.7313 
\end{tabular}
\end{center}
\end{table*}

The specific model selection process for the Coyote Universe runs is
described at length in \cite{coyote2}. It is based on Symmetric Latin
Hypercube (SLH) sampling \citep{slh}. Following the SLH strategy
guarantees good coverage of the parameter hypercube. In our specific
case we chose an SLH design that has space filling properties in the
case of two-dimensional projections in parameter space. In other
words, if any two parameters are shown in a plane, the plane will be
well covered by simulation points. An extensive discussion of optimal 
design choices is given in \cite{coyote2}. The interested reader is 
referred to that paper for details.

\begin{table*}
\begin{center} 
  \caption{Box sizes, particle numbers, force resolution, and
    corresponding values for shot noise limit, Nyquist frequency, and
    mass resolution. \label{tab:nest}}
\begin{tabular}{ccccccccc}
Length & $N_p^3$ & Number of & Force res. & Shot noise        &
$k_{\rm Ny}$     & $m_p$ & Scale factors\\ 
$[$Mpc$]$    &\hfill   & Realizations        & [kpc]           &
$[$Mpc$^{3}]$ &$[$Mpc$^{-1}]$  & $[$M$_\odot]$ & \hfill\\ 
\hline
1300 & $512^3$  &  16 &1270 & 2.05 & 2.48 & $5.7\cdot 10^{11}\omega_m $ &0.2 -- 1.0\\
1300 & $1024^3$  &  3 &635 & 2.05 & 2.48 & $5.7\cdot 10^{11}\omega_m $ &0.2 -- 1.0\\
1300 & $1024^3$  &  1 &50 & 2.05 & 2.48 & $5.7\cdot 10^{11}\omega_m $  &0.2 -- 1.0\\
365   & $512^3$    &   2   & 10 & 0.36 & 4.41 & $1.0\cdot 10^{11}\omega_m $ &0.4 -- 1.0\\
180   & $512^3$    &   3   & 10 & 0.04& 8.94 & $1.2\cdot 10^{10}\omega_m $ &0.2 -- 0.6\\
90     & $512^3$    &    4& 10 & 0.005& 17.87 & $1.5\cdot 10^{9}\omega_m $ &0.2 -- 0.33\\
\end{tabular}
\end{center}
\end{table*}

The emulator developed here is valid over the redshift range
$z=\{0,4\}$ and the $k$ range extends out to $k=8.6$~Mpc$^{-1}$. In
units of $h$Mpc$^{-1}$ this covers $k$ ranges between $k\sim
10~h$Mpc$^{-1}$ and $k\sim 15~h$Mpc$^{-1}$, depending on the
particular cosmological model. In order to enable a smooth
interpolation between redshifts, we store results at 11 outputs:
\begin{eqnarray}
a=&&\{1.0,~0.9,~0.8,~0.7,~0.6,~0.5,0.4,\nonumber\\
&&~~0.3333,~0.2857,~0.25,~0.2\}.
\end{eqnarray}
We use different box sizes to cover different ranges in $k$ and
redshift space. A summary of the simulation sizes is given in
Table~\ref{tab:nest}. The largest set of simulations is from the
original Coyote Universe suite as described in \cite{coyote3}. This
set of runs (all carried out in a 1300~Mpc box) consists of -- for
each cosmological model -- 16 realizations with 512$^3$ particles
using a particle mesh (PM) code with a 1024$^3$ grid, 4 PM
realizations with 1024$^3$ particles run on a 2048$^3$ grid, and one
high resolution {\sc Gadget-2} TreePM run \citep{gadget2} with
1024$^3$ particles. Many detailed tests of the high resolution
simulations including initial condition and resolution tests are described
in \cite{coyote1}. 
In addition, for the results reported here we run,
for each model, simulations with 512$^3$ particles and 10~kpc force
resolution with {\sc Gadget-2} in a 365~Mpc box (one realization per
model), a 180~Mpc box (three realizations per model), and a 90~Mpc box
(4 realizations per model). The initial conditions for these smaller volume simulations
are set up in a similar way to the large volume simulations: they are started at $z_{\rm in}=200$ 
using the Zel'dovich approximation.
As we explain below, we do not run the
90~Mpc box to $z=0$ but stop at $z=2$.  We will discuss the reasons
for our specific choices and how we match the different boxes in the
following section.

\subsection{Nested Simulations}
\label{sec:nested}

The extension of the original emulator beyond $k=1~$Mpc$^{-1}$ and
$z=1$ at high accuracy is nontrivial due to two limiting factors: the
spatial Nyquist frequency,
\begin{equation}
k_{\rm Ny}=\frac{\pi N_p}{L},
\end{equation}
setting the largest viable $k$-value, and the particle shot noise limit:
\begin{equation}
P_{\rm shot}(k)=\left(\frac{L}{N_p}\right)^3,
\end{equation}
restricting the lowest amplitudes at which the power spectrum can be
measured accurately. These issues have been known for a long time
(see, e.g,  \citealt{baugh95} for an early discussion). With the recent
efforts to obtain very accurate results for the absolute measurement
of the power spectrum out to small scales, it is worthwhile to briefly
revisit the essential arguments.

Both the Nyquist limit and the shot noise level depend on the mass
resolution and the size of the simulation volume and in both cases
reasonable results can only be obtained for a large number of
particles. Computational limits imply a necessarily finite number of
particles; the obvious option is to shrink the simulation volume. But
this option has its own pitfalls -- lack of long-range power and
increased sampling variance, for example -- and will break down at
some point as discussed quantitatively below.

The shot noise effect is particularly annoying at early times; at late
times, once the power spectrum has risen substantially above the shot
noise level, it becomes much less problematic.  The Nyquist sampling
issue, which essentially sets the dynamic range at which the initial
condition can be produced, is worse for larger boxes with the same
particle number.  In addition, the impact of both problems depends on
$\sigma_8$.  For two simulations that differ only in their value of
$\sigma_8$, the one with the higher $\sigma_8$ will contain more
nonlinear structure and growth at the same redshift -- this means that
the shot noise problem is worse for low $\sigma_8$ runs while the
issues with small volumes at lower redshift are more severe for high
$\sigma_8$ models.

\begin{table*}
\begin{center} 
\caption{Assembly 1, $\sigma_8> 0.8$\label{tab:assem1}}
\begin{tabular}{lccccc}
$a$ &RPT& 1300~Mpc & 365~Mpc & 180~Mpc &90~Mpc\\
\hline
$0.2$ & $k\le 0.1$ &$ 0.1<k\le 0.4 $&  &$0.4\le k\le 1.0 $ &$1\le k\le
8.6$  \\ 
$0.25$ & $k\le 0.1 $&$0.1< k\le 0.4 $  &\hfill  &$0.4\le k\le 1.0 $
&$1\le k\le 8.6$  \\ 
$0.2857$ & $k\le 0.1$ &$ 0.1<k\le 0.4 $  &\hfill  &$0.4\le k\le 1.0 $
&$1\le k\le 8.6$  \\ 
$0.3333$ & $k\le 0.1$ &$0.1< k\le 0.4 $ &\hfill  &$0.4\le k\le 1.0
$&$1\le k\le 8.6$  \\ 
$0.4$ & $k\le 0.1$ & $0.03<k\le 0.4 $ &$0.4 <k\le 8.6$  &\hfill
&\hfill  \\ 
$0.5$ & $k\le 0.03$ & $0.03<k\le 1.0 $ &$1.0 <k\le 8.6$  &\hfill
&\hfill  \\ 
$0.6$ & $k\le 0.03$ & $0.03<k\le 1.0 $ &$1.0 <k\le 8.6$  &\hfill
&\hfill  \\ 
$0.7$ & $k\le 0.03$ & $0.03<k\le 1.0 $ &$1.0 <k\le 8.6$  &\hfill
&\hfill  \\ 
$0.8$ & $k\le 0.03$ & $0.03<k\le 1.0 $ &$1.0 <k\le 8.6$  &\hfill
&\hfill  \\ 
$0.9$ & $k\le 0.03$ & $0.03<k\le 1.0 $ &$1.0 <k\le 8.6$  &\hfill
&\hfill  \\ 
 $1.0$ & $k\le 0.03$ & $0.03<k\le 1.0 $ &$1.0 <k\le 8.6$  &\hfill
&\hfill   
\end{tabular}
\end{center}
%
\begin{center} 
\caption{Assembly 2, $0.7\le\sigma_8\le 0.8$\label{tab:assem2}}
\begin{tabular}{lccccc}
$a$ &RPT& 1300~Mpc & 365~Mpc & 180~Mpc &90~Mpc\\
\hline
$0.2$ & $k\le 0.1$ &$ 0.1<k\le 0.4 $&  &$0.4\le k\le 1.0 $ &$1\le k\le
8.6$  \\ 
$0.25$ & $k\le 0.1 $&$0.1< k\le 0.4 $  &\hfill  &$0.4\le k\le 1.0 $
&$1\le k\le 8.6$  \\ 
$0.2857$ & $k\le 0.1$ &$ 0.1<k\le 0.4 $  &\hfill  &$0.4\le k\le 1.0 $
&$1\le k\le 8.6$  \\ 
$0.3333$ & $k\le 0.1$ &$0.1< k\le 0.4 $ &\hfill  &$0.4\le k\le 1.0
$&$1\le k\le 8.6$  \\ 
$0.4$ & $k\le 0.1$ & $0.03<k\le 0.4 $ &\hfill  &$0.4 <k\le 8.6$
&\hfill  \\ 
$0.5$ & $k\le 0.03$ & $0.03<k\le 1.0 $ &$1.0 <k\le 8.6$  &\hfill
&\hfill  \\ 
$0.6$ & $k\le 0.03$ & $0.03<k\le 1.0 $ &$1.0 <k\le 8.6$  &\hfill
&\hfill  \\  
$0.7$ & $k\le 0.03$ & $0.03<k\le 1.0 $ &$1.0 <k\le 8.6$  &\hfill
&\hfill  \\ 
$0.8$ & $k\le 0.03$ & $0.03<k\le 1.0 $ &$1.0 <k\le 8.6$  &\hfill
&\hfill  \\ 
$0.9$ & $k\le 0.03$ & $0.03<k\le 1.0 $ &$1.0 <k\le 8.6$  &\hfill
&\hfill  \\  
$1.0$ & $k\le 0.03$ & $0.03<k\le 1.0 $ &$1.0 <k\le 8.6$  &\hfill
&\hfill   
\end{tabular}
\end{center}
%
\begin{center} 
\caption{Assembly 3, $\sigma_8< 0.7$\label{tab:assem3}}
\begin{tabular}{lccccc}
$a$ &RPT& 1300~Mpc & 365~Mpc & 180~Mpc &90~Mpc\\
\hline
$0.2$ & $k\le 0.1$ &$ 0.1<k\le 0.4 $&  &$0.4\le k\le 1.0 $ &$1\le k\le
8.6$  \\ 
$0.25$ & $k\le 0.1 $&$0.1< k\le 0.4 $  &\hfill  &$0.4\le k\le 1.0 $
&$1\le k\le 8.6$  \\ 
$0.2857$ & $k\le 0.1$ &$ 0.1<k\le 0.4 $  &\hfill  &$0.4\le k\le 1.0 $
&$1\le k\le 8.6$  \\ 
$0.3333$ & $k\le 0.1$ &$0.1< k\le 0.4 $ &\hfill  &$0.4\le k\le 1.0
$&$1\le k\le 8.6$  \\ 
$0.4$ & $k\le 0.1$ & $0.03<k\le 0.4 $ &\hfill  &$0.4 <k\le 8.6$
&\hfill  \\  

$0.5$ & $k\le 0.03$ & $0.03<k\le 1.0 $ &\hfill &$1.0 <k\le 8.6$
&\hfill  \\ 
$0.6$ & $k\le 0.03$ & $0.03<k\le 1.0 $ &\hfill &$1.0 <k\le 8.6$
&\hfill  \\ 

$0.7$ & $k\le 0.03$ & $0.03<k\le 1.0 $ &$1.0 <k\le 8.6$  &\hfill
&\hfill  \\ 
$0.8$ & $k\le 0.03$ & $0.03<k\le 1.0 $ &$1.0 <k\le 8.6$  &\hfill
&\hfill  \\ 
$0.9$ & $k\le 0.03$ & $0.03<k\le 1.0 $ &$1.0 <k\le 8.6$  &\hfill
&\hfill  \\ 
$1.0$ & $k\le 0.03$ & $0.03<k\le 1.0 $ &$1.0 <k\le 8.6$  &\hfill
&\hfill   
\end{tabular}
\end{center}
\end{table*}

At the current levels of available computational power, it is
impossible to carry out a brute force approach to obtain a sub-percent
accurate answer for the power spectrum at the desired $k$-values. As
discussed at length in  \cite{coyote1}, the maximum value is set by
$k_{\rm Ny}/2$. At the same time, in order to ensure that the largest
modes in the box at $z=0$ are linear to high accuracy, a linear box
size of 2000~Mpc would be a good choice. To reach a wavenumber value
of 10~Mpc$^{-1}$ would imply a particle loading of $N_p^3\sim
12700^3\sim 2$ trillion particles. While it is possible to carry out a
few such simulations \citep{habib12}, a large suite of them is clearly
currently out of reach.

This notional set-up would lead to a shot noise level of $\sim 4\cdot
10^{-4}$. The model with the lowest $\sigma_8$ in our simulation set
is M019 with $\sigma_8=0.6159$. The amplitude of the power spectrum at
$z=4$ is $P(k=10,z=4)=0.025$~Mpc$^{3}$, translating to $\sim 60$ times
above the shot noise level and is, hence, safe. If we wanted to
decrease the number of particles by choosing a smaller volume,
1000~Mpc would be barely large enough (for more detailed discussions
of finite volume effects see e.g. \citealt{coyote1}). In this case, $\sim$ 250
billion particles would be needed to push out the Nyquist limit far
enough. This would lead to a shot noise level of $\sim 4\cdot
10^{-3}$, dangerously only a factor of 6.25 below the amplitude of the
power spectrum at $k=10$~Mpc$^{-1}$ and $z=4$. Moreover, even this
simulation size constitutes a currently prohibitive expense if a full
suite of simulations has to be performed.

Due to these obstacles, a strategy for mixing and matching simulation
sizes -- differently for different redshifts -- needs to be
employed. Smaller simulation volumes are required for high redshift
results while larger volumes will be used for lower
redshifts. Although such a procedure is effective as we show below,
because of the uncertainties induced by having to sew together
multiple simulation results, and because the number of cosmological
models used is still the same as in the original set, sub-percent
accurate power spectra are not obtained over the new, much wider
dynamic range in wavenumber and redshift. At the small spatial scales
that the current emulator goes to, however, current uncertainties in
characterizing baryonic effects clearly outweigh its
inaccuracies. Therefore, over the extended dynamic range, the
constraints on the accuracy of the dark matter power spectrum are not
uniform, and not as stringent near the limits of the range probed in
wavenumber and in redshift.

We quantify below some of the errors due to the fact that the power
spectra at all $k$ and $z$ do not arise from a single overarching
simulation. Over its full range, and across all cosmological models,
the emulator is nevertheless accurate to better than $5\%$. We now
describe the details of our matching strategy for the power spectra at
each redshift. We store results at 11 scale factors
($a=0.2,0.25,0.2857,0.3333,0.4,0.5,0.6,0.7,0.8,0.9,1.0$, see also
Tables~\ref{tab:assem1} -- \ref{tab:assem3}) and then interpolate
between those outputs to obtain results at any scale factor in
between.

\begin{figure}[t]
\begin{center}
\includegraphics[width=80mm]{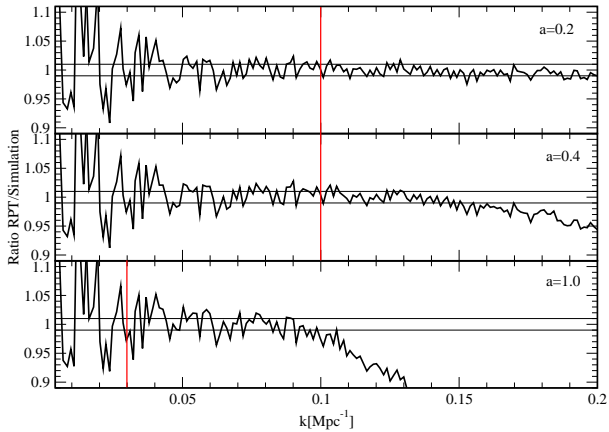}
\end{center}
\caption{Comparison of the simulations with renormalized perturbation
  theory for M037 (averaged over 20 realizations) at three
  redshifts. The black horizontal lines show the 1\% error limits. The
  red lines show the matching points we choose for connecting
  perturbation theory to the simulation results. For $1.0\le a \le
  0.5$ the matching value is at $k=0.03~$Mpc$^{-1}$ and for $0.4\le
  a\le 0.2$ it is at $k=0.1~$Mpc$^{-1}$.}
\label{pert_test}
\end{figure}

For the largest scales, we use renormalized perturbation theory [RPT,
see~\cite{crocce1} and \cite{crocce2} for details on the underlying
idea] using the Copter code \citep{copter}; the code has been modified
to allow for $w\ne-1$.  We start by breaking up the models into three
groups, depending on the power spectrum normalization, $\sigma_8$:
\begin{eqnarray}
\sigma_8 < 0.7,~{\rm Assembly~1},\\
0.7\le\sigma_8\le0.8,~{\rm Assembly~2},\\
\sigma_8 > 0.8,~{\rm Assembly~3}.
\end{eqnarray}
We use RPT for all three cases in the same way up to the following $k$
values: 
\begin{eqnarray}
&&1.0 \le a\le 0.5:~k=0.03~{\rm Mpc}^{-1}\nonumber\\
&&0.4 \le a\le 0.2:~k=0.1~{\rm Mpc}^{-1}.\nonumber
\end{eqnarray}
(As described above, we do not store any outputs between $a=0.4$ and $a=0.5$.) 
The first cut is the same as was used in \cite{coyote3}, while the
second is more aggressive. Since the power spectrum is more linear at
higher redshifts, perturbation theory will be valid out to higher
$k$. We exhibit the accuracy of RPT at the matching scales for the
model with the highest $\sigma_8$, M037 ($\sigma_8=0.9$) -- the most
difficult case -- in Fig.~\ref{pert_test}. The ratio of RPT is shown
with respect to the simulation output (the average from our 20
realizations with $L=1300$~Mpc) at three different redshifts, $z=0,
1.5, 4$. The red vertical line shows our matching point for
perturbation theory. In all cases, the error is below 1\%, in very
good agreement with the findings of \cite{copter}.

Next, we must determine the maximum $k$ values out to which to use the
results from each of the different volumes. The individual matching
strategies for the three assemblies depending on $\sigma_8$ are
summarized in Tables~\ref{tab:assem1} -- \ref{tab:assem3}. As
mentioned above, the matching point depends on the Nyquist wavenumber
and, at higher redshifts, the shot noise level. As previously
established in  \cite{coyote1} the results from the 1300~Mpc volume
hold at the sub-percent level accuracy out to $k\sim 1~$Mpc$^{-1}$
between $0\le z\le 1$ or $0.5\le a\le 1.0$. Therefore, the power
spectra for all three assemblies in this range are taken from these
simulations (the simulations are also particularly valuable because of
the large numbers of realizations for each of the boxes).

At higher redshifts, the shot noise level of the earlier simulations
becomes too high and we can use them only to a lower $k$-value,
consequently, in all three assemblies, for $0.2\le a\le 0.4$ we cut
off the power spectra from the 1300~Mpc boxes at
$k=0.4$~Mpc$^{-1}$. For $a \ge 0.5$, the upper $k$ value is taken to
be 1.0~Mpc$^{-1}$. The following step is to match the results from the
smaller boxes at these cutoff points. Shot noise level considerations
play the dominant role in choosing the matching points for the power
spectra from the different simulations.

In the case of Assembly~1 (high $\sigma_8$), the second-largest box
(365~Mpc) is used to cover the remaining full range in wavenumber from
$0.4\le a \le 1$. The next two smaller boxes of size 180~Mpc and
90~Mpc are used to fill in the higher $k$ range as shown in
Table~\ref{tab:assem1}. The results of analogous strategies for
Assembly~2 (medium $\sigma_8$) and Assembly~3 (low $\sigma_8$) are
given in Table~\ref{tab:assem1} and Table~\ref{tab:assem3}.

\begin{figure*}[t]
\begin{center}
\includegraphics[width=170mm]{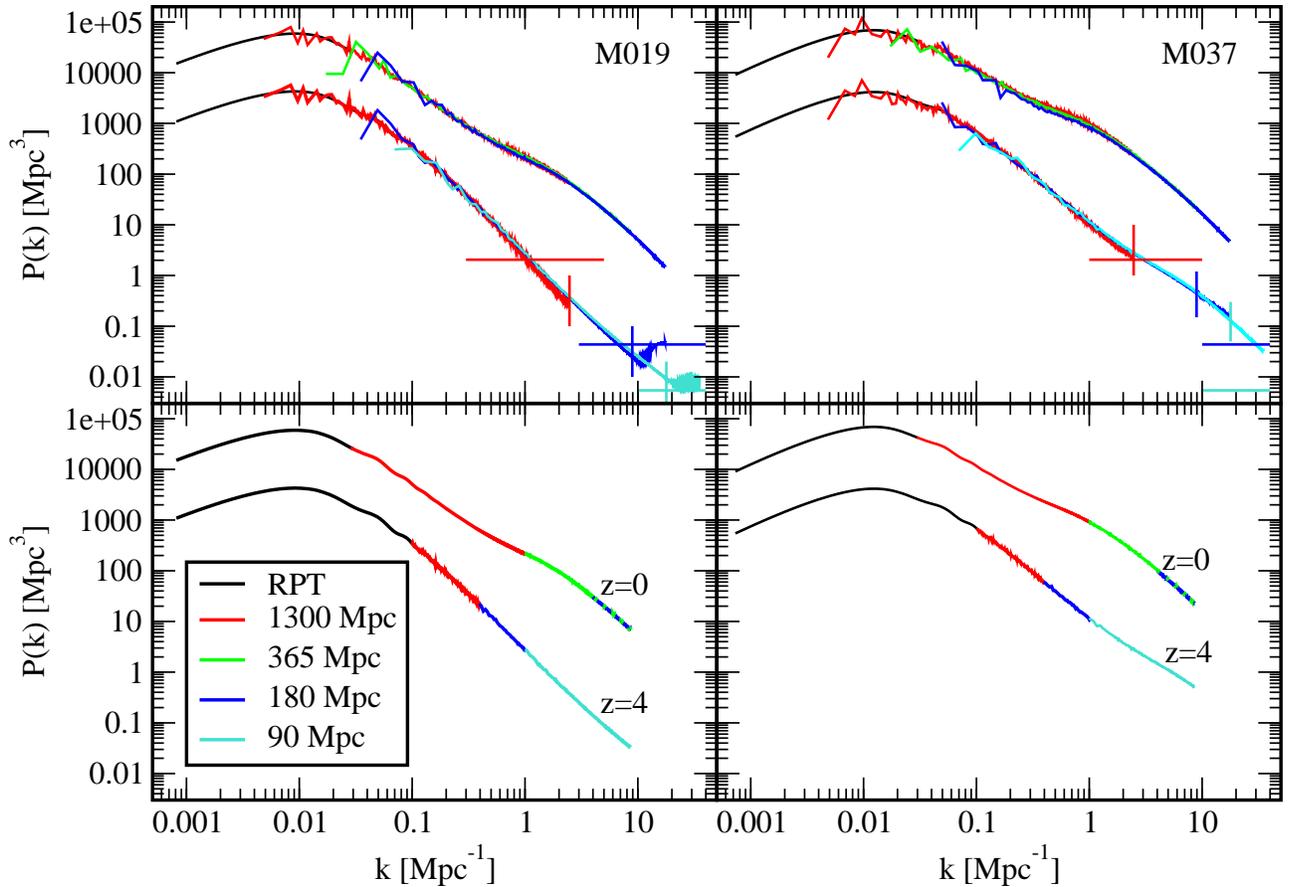}
\end{center}
\caption{Two examples of how nested simulation volumes are used to
  cover a large range in $k$. Models shown have the most extreme
  values of $\sigma_8$, M019 with $\sigma_8=0.6159$ (right column) and
  M037 with $\sigma_8=0.9$ (left column). The upper row shows the full
  power spectra from the different simulations as well as the
  corresponding Nyquist limit (vertical line) and the shot noise limit
  (horizontal line). The lower row shows the different power spectra
  matched up (without any additional smoothing) at the $k$-values
  described in Tables~\ref{tab:assem1} -- \ref{tab:assem3}. The power
  spectra results are shown at $z=0$ and $z=4$.  In all cases the
  matching leads to a relatively smooth power spectrum covering the
  full $k$ range of interest. In the lower panel, at $z=0$, we show
  the good agreement between the 365~Mpc and 180~Mpc boxes at high $k$
  by overlaying the two power spectra (see text for details). Note
  that in the lower plots at $z=0$ we used the original emulator
  predictions for the intermediate $k$ range shown in red. }
\label{match}
\end{figure*}

Figure~\ref{match} shows the matched power spectra for two example
models, M019 and M037, representing the lowest and highest $\sigma_8$
in our model selection. The upper panels show the full power spectra
from all boxes while the lower panels show the results with cut-offs
applied at the appropriate $k$-values. The results in the lower panel
demonstrate that the approach of using nested volumes is
qualitatively satisfactory. In the next two sub-sections we discuss
the main errors involved in the matching procedure, (i) finite mass
resolution which leads us to push beyond $k_{\rm Ny}/2$ for the small
boxes, and (ii) finite volume effects.

\subsection{Finite mass resolution effects}

In \cite{coyote1} the effects of finite mass resolution on power
spectrum estimation were discussed in some detail. Two effects
investigated there act in opposite directions. The first is sampling
noise, i.e., the effect on computing a power spectrum from a density
field computed from a different number of particles. For this, a
particle distribution at $z=0$ was down-sampled by factors of 8, 64,
and 512 and the power spectra re-measured. The effect of the
insufficient sampling of the density field led to an enhancement of
the power spectrum larger than 1\% beyond $k_{\rm Ny}/2$. The second
is the effect of lower particle sampling in the initial
conditions. This leads to a suppression of the power spectrum since
the initial conditions now lack small-scale power. This effect
decreases for simulations with larger Nyquist frequency since the
number of modes in the box at small scales increases. In the test in
\cite{coyote1} it was shown that a simulation with $k_{\rm
  Ny}$=0.859~$h$Mpc$^{-1}$ has lost more than 10\% of power at $k_{\rm
  Ny}/2$, while for $k_{\rm Ny}$=3.44~$h$Mpc$^{-1}$ it is well below
1\% at $k_{\rm Ny}/2$.

In the current paper, we use the large volume simulations (1300~Mpc)
only up to $k\le 1~$Mpc$^{-1}$ at low redshift and $k\le
0.4~$Mpc$^{-1}$ at high redshift, both values being well below $k_{\rm
  Ny}/2$. For the smallest volumes (90~Mpc) we also do not use any
results beyond $k_{\rm Ny}/2$, simply because for the smallest boxes
the Nyquist limit is much larger than $k\sim 10~$Mpc$^{-1}$.  For the
intermediate boxes (180~Mpc and 365~Mpc) we choose to be more
aggressive.  For the lowest redshifts we push the 365~Mpc limit to
twice the Nyquist criterion and for intermediate redshifts we use the
180~Mpc box out to the Nyquist frequency. In principle, we could have
avoided pushing the 365~Mpc box beyond the Nyquist limit by using the
180~Mpc box at large $k$ values instead. There are two reasons why we
chose not to do this.

First, the comparison of inaccuracies due to finite box effects versus
going beyond the Nyquist limit suggests that the finite box errors
dominate. The quasi-linear regime in the small box simulations close
to the final time step at which they are used is not accurate because
of the missing large-scale power. In addition, the scatter in the
overall amplitude is larger due to the small box size as discussed in
the next sub-section. These two errors combine to outweigh the
inaccuracy due to the Nyquist limit. This point is illustrated in
Fig.~\ref{match}. In the right upper panel (M037), at $z=0$, the green
curve shows the result for the 365~Mpc box while the blue curve shows
the result for the 180~Mpc box. It is apparent that the small box
clearly does not capture the nonlinear turn-over behavior beyond
$k\sim 0.1$~Mpc$^{-1}$ very well. On the other hand, both results are
in close agreement at large $k$, so the accuracy of the 365~Mpc box
results is still good at that point.

Second, the matching procedure described in Section~\ref{sec:smooth}
for melding the different power spectra pieces induces another small
inaccuracy. This is again mainly due to finite box size effects, as
the amplitude has to be adjusted to enable a matching of the boxes in
the high $k$ region. It is therefore desirable to minimize the number
of matching points, and hence to take the results from the 365~Mpc
boxes out to higher $k$ values, rather than to introduce another
matching point at intermediate scales.  In order to verify that using
the 365~Mpc box at higher $k$ does not introduce a major error, we
also checked that the difference at high $k$ between the 365~Mpc box
and the 180~Mpc box is small, as can be seen in Fig.~\ref{match}. For
M037 (high $\sigma_8$) we find differences of 2\% and less beyond
$k=6$~Mpc$^{-1}$ and for M019 (low $\sigma_8$) we find differences
well below 1\% in the high $k$ range.

To summarize this discussion, for the high redshift results, we do not
use simulations beyond $k_{\rm Ny}/2$ (see
Tables~\ref{tab:assem1}-\ref{tab:assem3} for more details) and
therefore estimate the error throughout the $k$ range as being well
below 1\%, based on the test results from \cite{coyote1}.  For the low
redshift cases, we use some results that go beyond the Nyquist
limit. By comparing to the smaller boxes we estimate that the error
does not exceed 2\%.

\subsection{Finite box size effects}

\begin{figure}[t]
\begin{center}
\includegraphics[width=90mm]{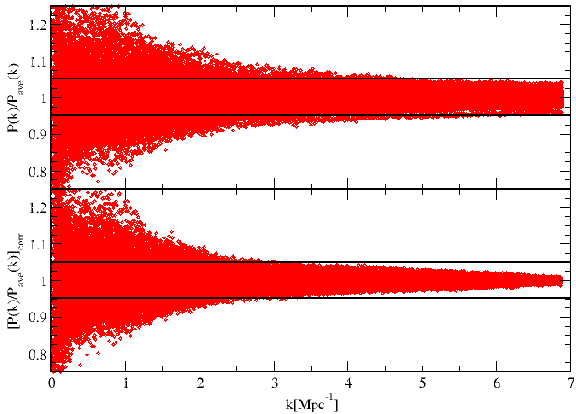}
\end{center}
\caption{Variations in power spectra due to different realizations
  from 127 simulations, each for a 325~Mpc box size. The upper panel
  shows the ratio of each of the 127 results at $z=0$ with respect to
  the average power spectrum. The lower panel shows an attempt to
  correct the results -- here we adjust the amplitude of each power
  spectrum so that it matches the average power spectrum at the
  largest $k$-value. While at large $k$ this reduces the error
  somewhat, the procedure is not satisfactory overall and we do not
  use it in the final results. The scatter is roughly at the few
  percent level. }
\label{127sim}
\end{figure}

As outlined in the Introduction, ultimately one would want to carry
out simulations in large box sizes, 1-2~Gpc, but this is currently
impractical for a large suite of runs, each with sufficient mass
resolution. Smaller simulation volumes have two drawbacks: (i) At
redshifts close to $z=0$ the undersampling of large-scale power may be
a problem (the precise extent depending on $\sigma_8$ and the box
size). In smaller boxes, while the largest-scale modes might appear to
be linear, their amplitude is actually suppressed by missing nonlinear
power (in a large volume run, the same modes would have higher
amplitudes). (ii) The realization scatter in small volumes is much
larger due to the smaller number of large scale modes. The first of
these points was considered in depth in \cite{coyote1}. There, we
demonstrated the suppression of the power spectrum due to finite
volume effects by comparing the average of 4 realizations in a
2000~$h^{-1}$Mpc box, 8 realizations in a 936~$h^{-1}$Mpc box, and 127
realizations in a a 234~$h^{-1}$Mpc box. Figure~6 in \cite{coyote1}
shows the suppression at mildly nonlinear scales. The realization
scatter was not a significant issue in that paper since the final
simulation volumes were large (1300~Mpc) and we averaged over 20
realizations.

\begin{figure}[t]
\begin{center}
\includegraphics[width=60mm,angle=270]{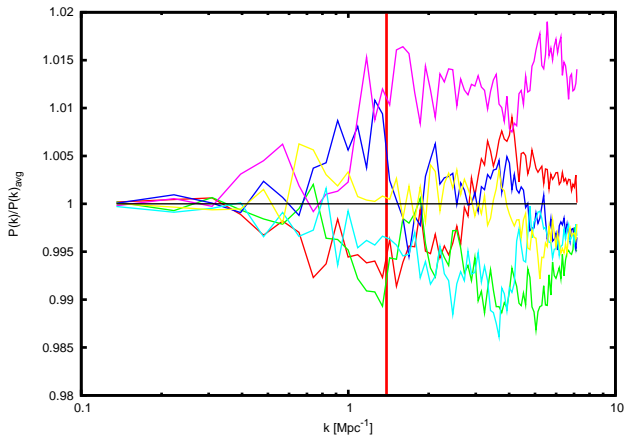}
\end{center}
\caption{Test of realization scatter on small scales. Shown are the
  results from six simulations that have the same realization on large
  scales but different realizations on small scales divided by the average 
  of all six power spectra. The red line
  indicates the transition between the two regimes.  Note that some of
  the realization noise from the small scales leaks back to larger
  scales. }
\label{real}
\end{figure}

In the current paper, the realization scatter is of concern since we
do resort to using small volumes. The mildly nonlinear regime is still
very accurate since it is covered by large volume simulations, so the
key question is to understand the high-$k$ behavior. In order to
investigate this effect, we carry out two tests. First, we use the
same 127 simulations used in \cite{coyote1} to measure the overall
dispersion in the power spectra between different realizations. These
simulations were carried out in 325~Mpc volumes with a PM code almost
matching the small simulation volumes used here. Since we are only
interested in the relative effect, these lower resolution simulations
are sufficient. The results are encapsulated in Fig.~\ref{127sim}. The
upper panel shows the ratio of the average of the 127 power spectra
with respect to each individual spectrum; in the lower panel, we
attempt to correct the run-to-run scatter by simply adjusting the
amplitude of each power spectrum to the average value at the highest
measured $k$ (which is well-sampled in each simulation). This
procedure improves the scatter somewhat but is rather uncontrolled and
we decided not to use it. Nevertheless, the result is somewhat
interesting.  Overall, the run-to-run scatter at small scales is up to
about 5\%, at the matching scale of $k\sim 1$~Mpc$^{-1}$, and up to
10\% for the most extreme outliers. In order to reduce this problem,
we carry out at least two realizations for each model, in some case up
to four to obtain a realization in which the matching to the 1300~Mpc
box is as close as possible. Since we match the even smaller volumes
(180~Mpc and 90~Mpc) at larger $k$ values, the problem there is not
quite as severe. Nevertheless, even in those cases we run up to four
realizations per model to provide an accurate match to the larger
boxes.

As a second check we carry out six simulations of the same cosmology
each with box length $L = 365~$Mpc and $N_p^3 = 512^3$ particles,
which are chosen to exactly match the size and resolution of the
smaller Coyote runs.  The initial particle distribution is set with
the same realization at $z=211$ on large scales for all six
simulations, but on scales smaller than $k = 1.39$~Mpc$^{-1}$, we
allow the initial conditions to vary between runs by changing the
random seed for each simulation. These are then evolved using {\sc
  Gadget-2} and we show the ratio of power spectra at $z=0$ with
reference to a single simulation in the set in Fig.~\ref{real}.  The
red line marks where the initial power spectrum differs between the
realizations. Notice that the scatter in the power spectrum is less
than $\sim3\%$ and appears unbiased. This gives an estimate of the
amount of mismatch between the 1300~Mpc and 365~Mpc Coyote simulations
that we can expect after matching their power spectra. There is also a
small amount of leakage of power from small to large scale modes,
since the power spectra only match at $k \ll 1.39$ Mpc$^{-1}$, where
they were seeded identically in the initial conditions.

In terms of the effects of the missing low-$k$ power, at z=0, for a
$\sim100$~Mpc box, the {\em rms} amplitude of the DC mode is at the
level of $\sim 10$\%, falling to a few percent at $\sim 300$~Mpc (see,
e.g., \citealt{gnedin11}).  These numbers are sufficiently small --
given the level of accuracy we are aiming to attain here -- that a
more sophisticated correction procedure is not required. As previously
mentioned, box sizes in the 1-2~Gpc range are sufficiently large to
render these effects sub-dominant when targeting accuracy levels of
$\sim 1$\%~\cite{coyote1}. As more computer power becomes available,
construction of the next generation of emulators will profit from the
increased volume and better mass resolution, both of which are
essential to improve the accuracy at higher wavenumbers.

To summarize, finite box size effects are clearly an important
issue. Some of the problems such as realization scatter can be
overcome by generating a large number of realizations (though this is
expensive) but some small inaccuracies in the power spectrum will be
unavoidable until larger volume, high-particle loading simulations can
be carried out.

\section{Hubble Parameter Extension}
\label{sec:hubble}

In our original work, the value for the Hubble parameter was
automatically determined from the other five cosmological parameters
to be the best-fit CMB value. To allow for more flexibility,
especially keeping in mind possible tensions in different
observational values of $h$, we now aim to allow it to vary within the
range that is covered by the original Coyote Universe runs, i.e.,
$0.55\le h \le 0.85$. This means that the sampled parameter space must
be suitably extended.

A first idea might be to use the existing 37 models (all of which have
different values of $h$) and rebuild the emulator keeping the Hubble
parameter free. This naive approach is inadequate. Figure~6 in
\cite{coyote2} (seventh row) shows the distribution of $h$ values with
respect to the other parameters. From that figure it is clear that the
parameter space for $h$ is not well covered if we restrict ourselves
to the 37 available models. The design has large holes, particularly
for high values of $h$. Our tests confirmed that this sub-optimal
design does not lead to an accurate emulator. We note that this result
is in fact a successful demonstration of our overall approach, since
it shows that the optimality of the sampling design is indeed a
crucial factor.

\begin{figure}
\begin{center}
\includegraphics[width=80mm]{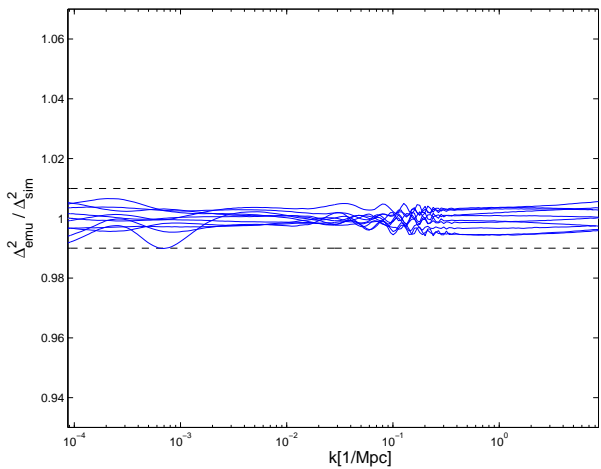}
\end{center}
\caption{Accuracy of the linear power spectrum emulator of the full
  six-dimensional parameter space. Shown is the ratio of the emulator
  to the linear power spectra for ten extra models not used for
  constructing the emulator itself. The error overall is within
  1\%. Note that we only included information on the Hubble parameter
  $h$ on large scales, with an upper cutoff at $k=0.03~$Mpc$^{-1}$. }
\label{hubble_test}
\end{figure}

The obvious alternate approach to include $h$ as a new parameter is to
generate a new design over six parameters and increase the number of
models to obtain sufficient accuracy.  This is obviously undesirable
for just one additional parameter since it would add a large
computational cost and would not make use of the already available
simulations. We therefore choose a quite different path.

The new idea implemented here is to exploit a certain flexibility in
constructing emulators; this flexibility relates to the fact that
emulators can be constructed by incorporating results from multiple
models across different scales, i.e., where the result for each model
does not necessarily have information available across the complete
set of scales.

We proceed by generating 100 new predictions over a six-dimensional
parameter space $\theta=\{\omega_b,\omega_m,n_s,h,w,\sigma_8\}$ with
RPT out to $k\le 0.1~$Mpc$^{-1}$ for high redshift ($z\ge 2$) and out
to $k\le 0.03~$Mpc$^{-1}$ for low redshift ($z< 2$). For larger
$k$-values we use the results from the original 37 simulations for
which we already have high-accuracy predictions in the nonlinear
regime. The idea behind this approach is that a significant amount of
information is readily available in the linear to mildly nonlinear
regime covered by RPT. In addition, the very accurate predictions at
low $k$, as provided by the 100 power spectra, helps to anchor the
power spectra correctly on large scales. The information on small
scales is then provided by the available 37 models that have been
fully simulated.

\begin{figure}[t]
\begin{center}
\includegraphics[width=90mm]{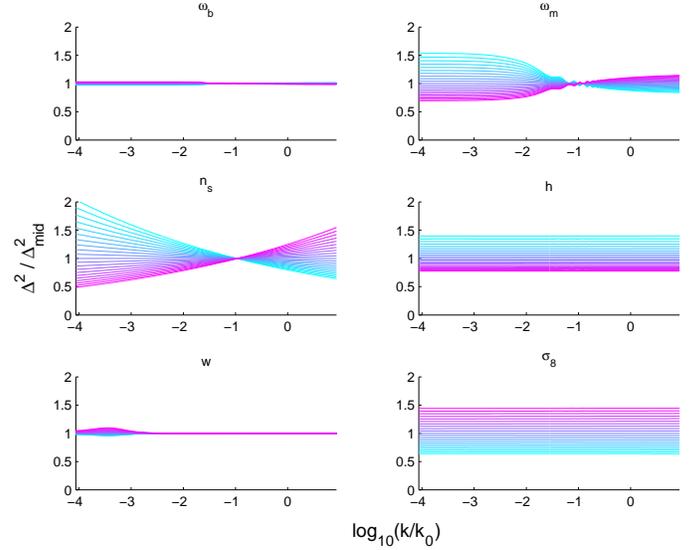}
\end{center}
\caption{Effects of the six different parameters within their
  respective prior ranges on the linear matter power spectrum.  Shown
  is the ratio with respect to a model where each parameter is fixed
  at its median value: $\theta=\{0.0225, 0.1375, 0.95, 0.7, -1.0,
  0.755\}$ and the sixth parameter (denoted at the top of each panel)
  is varied as a function of $k$ with $k_0$=1~Mpc$^{-1}$.  Light blue
  colors show results for small values of the sixth parameter, while
  pink colors show results for large values. The change of the power
  spectrum when varying the Hubble parameter $h$ is independent of
  scale, the reason why in the case of the linear power spectrum the
  addition of $h$ as a free parameter works well, even when using
  information only on very large scales.}
\label{sens}
\end{figure}

We modify the emulation procedure from \cite{coyote3} to account for
the inclusion of both `long' power spectra over the entire $k$ range
of interest and `short' power spectra in the low $k$ range
only. Recall that in \cite{coyote3}, the power spectra are modeled
using a basis representation:
\begin{equation}
\mathcal{P}(k;z;\theta)=
\sum_{i=1}^{n_\mathcal{P}}\phi_i(k;z)w_i(\theta),~~~\theta\in
[0,1]^{n_\mathcal{\theta}}, 
\end{equation}
where the $\phi_i(k;z)$ are the basis functions, the $w_i(\theta)$ are
the corresponding weights, and the $\theta$ represent the cosmological
parameters.  The basis vectors are constructed from principal
components and the weights are modeled as a Gaussian process. We again
use Gaussian processes to model the weights, but construct the basis
vectors slightly differently to better represent the variation in the
long and short power spectra.

Two features inform our selection of basis vectors: the inclusion of
RPT power spectra that cover only low $k$ and the combining of power
spectra from $N$-body simulations that use different box sizes.
First, we compute three principal component basis vectors using the 37
smoothed spectra resulting from combining $N$-body simulations of
different sizes.  Three basis vectors is sufficient to capture the
systematic variation at high $k$ and the shorter RPT spectra yield
well-behaved weights when projected on to these.  Second, we remove
the effects of the first three basis vectors and compute three
additional PC basis vectors on the residuals of the 37 smoothed
spectra, but only over the $k$ range of the original simulations.
This improves the resulting prediction of spectra over this $k$ range,
while avoiding extra variation at high $k$ resulting from the
combination.  Again, the residuals for the RPT power spectra project
on these basis vectors in a well-behaved manner.  Finally, we compute
three additional basis vectors on the residuals from all 137
simulations over the $k$ range covered by the RPT power spectra.  This
further improves the modeling over this low $k$ region.  For both sets
of basis vectors that cover only part of the $k$ range, the basis
vectors taper linearly to zero over their last 50 $k$ values to ensure
continuity.  This gives a total of nine basis vectors whose weights
are modeled with a Gaussian process.

In order to test the build-strategy that allows us to combine a
different number of simulations for small and large $k$, we carry out
a test with the linear power spectrum. We generate 100 predictions for
the linear power spectrum out to $k\le 0.03~$Mpc$^{-1}$ and in
addition 37 predictions following the original design over the full
$k$ range out to $k\le 10.0~$Mpc$^{-1}$.  We then build an emulator
over the full $k$ range allowing $h$ to vary. In order to test the
accuracy of the new emulator, we generate a set of 10 additional power
spectra over the full parameter range and compare the emulator
prediction with those exact results. Figure~\ref{hubble_test} shows
the ratio of the emulator prediction with the linear theory
answer. Overall, the accuracy is around 1\%, which is very
satisfactory.

With the new emulator in hand, we briefly study the dependence of the
power spectrum on the different cosmological parameters. For this we
generate a set of sensitivity plots, shown in Fig.~\ref{sens}. The
idea behind the sensitivity study is simple: we determine the power
spectrum for a model in the center of the parameter hypercube and then
vary one parameter at a time from its smallest to highest value. This
illustrates concisely the effect of each parameter on the power
spectrum on all scales studied. The Hubble parameter primarily shifts
the amplitude of the power spectrum up and down (right panel, middle
row), which explains why the simple addition scheme here works so
well.

We also produce a second emulator that keeps $h$ fixed.  This emulator
does not use the additional RPT spectra, but does include the results
from the smaller $N$-body simulations.  This emulator was constructed
in the usual manner, with $6$ PC basis functions computed from the 37
spectra over the whole $k$ range.

We will show in a future publication that the idea of combining
different numbers of models at different length scales is also very
useful in obtaining accurate predictions of the power spectrum on even
smaller scales.  As alluded to in the Introduction, computing power
spectra on such scales is computationally very expensive.  It is very
helpful if the number of models needed for this can be kept to a
minimum.

\section{Smooth Power Spectrum Generation}
\label{sec:smooth}

The power spectra from the $N$-body simulations are smoothed with
essentially the same process convolution model used in \cite{coyote3},
but with an important addition to account for vertical shifts where
power spectra from different simulations are pasted together.

In \cite{coyote3}, the simulation of cosmology $c$, resolution $s$,
and replicate $i$ produces a spectrum, $P^{c}_{s,i}$.  We model this
as a multivariate Gaussian variable with a known covariance $\Omega$,
and a smooth mean described by a process convolution
\citep{Higdon2002}.  A process convolution is constructed by
generating a latent stochastic process and smoothing it.  In
\cite{coyote3}, the spectra for each cosmology share a latent process,
$u^c$, modeled as a Brownian motion observed on a sparse grid.  These
latent processes are smoothed by a common smoothing matrix,
$K^{\sigma}$, made from Gaussian smoothing kernels whose kernel width
varies across the domain in order to account for nonstationarity.  The
resulting model has a probability density function
\begin{eqnarray} 
f(P^{c}_{s,i}) & \propto & \left| A_s \Omega A_s^{\prime}
\right|^{1/2} \\ & \times & \exp \left\{ -\frac{1}{2}
\left(P^{c}_{s,i} - A_s K^{\sigma} u^c \right)^{\prime} A_s \Omega
A_s^{\prime} \left(P^{c}_{s,i} - A_s K^{\sigma} u^c \right),
\right\}, \nonumber
\end{eqnarray}
where the matrix $A_s$ truncates the length of the spectrum depending
on the resolution.  The modeling details are given in \cite{coyote3}.

For the current work, we make a small addition to the mean function
for the spectrum for each simulation.  The spectra for the highest
resolution simulations are augmented with spectra from the smaller
boxes as described in the assembly tables.  At each matching point,
the spectra from each box may have a small vertical offset.  However,
we know that each simulation is approximating a smooth spectrum across
the $k$ range.  To account for this, we modify the mean structure for
the simulation model.  The mean structure is still built around a
process convolution model, but includes additional terms that move
each section of the simulated spectrum up or down.

Let $H^c$ be a matrix with a row for each $k$ value and three columns
(one for each of the small boxes -- no offset if estimated between the
perturbation theory and the original set of simulations).  Let the
cosmology index $c$ also index redshift (although unstated, this is
also true in  \citealt{coyote3}).  This matrix describes the matching.
At a particular $k$, if the simulated spectrum is represented by a
result from perturbation theory or the original simulations, the
matrix $H^c$ has a row of zeros.  If the simulated spectrum uses the
result from the 365~Mpc box, the row has a one in the first column.
If the simulated spectrum uses a result from the 180~Mpc box or the
90~Mpc box, the row has a one in the second or third column
respectively.  Each can have at most a single non-zero entry.  Let
$b^c$ be a set of three coefficients that represent the offset of the
simulated result from the unknown smooth mean.  We add these terms to
the model.  Let $\mathcal{P}^{c} = K^{\sigma} u^c + H^c b^c$, which
gives the density
\begin{eqnarray} 
f(P^{c}_{s,i}) & \propto & \left| A_s \Omega A_s^{\prime}
\right|^{1/2} \\ & \times & \exp \left\{ -\frac{1}{2}
\left(P^{c}_{s,i} - A_s \mathcal{P}^{c} \right)^{\prime} A_s \Omega
A_s^{\prime} \left(P^{c}_{s,i} - A_s \mathcal{P}^{c} \right)
\right\}. \nonumber
\end{eqnarray}
This model conveys the information that there is a smooth spectrum
represented by $K^{\sigma} u^c$ and that we observe a noisy version of
it that has vertical shifts over some ranges.

The offset parameters need a prior distribution to complete the
specification.  We give them a zero-mean Gaussian prior:
\begin{equation}
\pi(b) \propto \exp \left \{ -\frac{\lambda_b}{2} b'b \right\}.
\end{equation}
The parameter $\lambda_b$ is a precision parameter that we set equal
to the known precision of the simulated spectrum at the first matching
point.  Like the latent process $u^c$, the $b$ can be integrated out
of the posterior and need not be drawn as part of the Markov chain
Monte Carlo approach.

\section{Emulator for the Matter Power Spectrum}
\label{sec:emu}

\begin{figure}[t]
\begin{center}
\includegraphics[width=80mm]{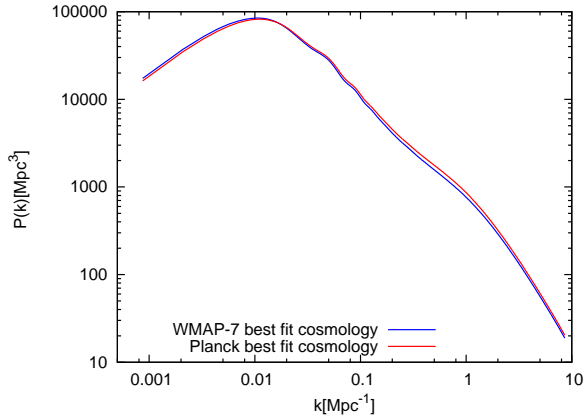}
\caption{Emulator predictions for the nonlinear power spectra for the
  best fit cosmologies for WMAP-7 and Planck (see text for details on
  the parameter choices) at $z=0$. Due to the higher values for
  $\omega_m$ and $\sigma_8$, the Planck results imply a noticeable
  change in the nonlinear regime.}
\label{fig:pks}
\end{center}
\end{figure}

The construction of the new power spectrum emulator now follows the
general methodology laid out in \cite{coyote2} and \cite{coyote3} with
the small modifications discussed above to combine power spectra over
different $k$ ranges. In this section we present some results from the
emulator and tests of its accuracy. To start, we show the predictions
of the emulator for the best-fit cosmology found by the WMAP-7 and
Planck surveys in Fig.~\ref{fig:pks}.  Following the notation in
Eq.~(\ref{eq:theta}) the parameters used are
\begin{eqnarray}\label{eq:params}
\theta_{\rm\small WMAP-7}&=&\left\{0.0225 , 0.13328, 0.97,  0.7, -1.0,
  0.81\right\},\\ 
\theta_{\rm\small Planck}&=&\left\{0.02225, 0.14026, 0.968, 0.6816 ,
  -1.0, 0.8284 \right\}\nonumber, 
\end{eqnarray}
as obtained from CMB measurements alone. While the spectra are close,
the Planck parameters lead to an enhancement of the power spectrum in
the nonlinear regime.

Below we describe a number of tests of the emulator accuracy. We find
that the emulator quality is better than 5\% over a large $k$ and $z$
range, even when including the new free parameter, $h$. We also
include a comparison against the latest improved version of Halofit as
implemented in CAMB (see \citealt{takahashi} for details).

\subsection{Power Spectrum Predictions}

As explained in Section~\ref{power} the emulator is built using
results from 37 cosmological models, specified in
Table~\ref{tab:basic}.  In addition to these models, we have generated
$P(k)$ results for a $\Lambda$CDM model (M000), which is close to the
current best-fit measurements. Simulations for this model were carried
out in exactly the same way as for the other models, using several
realizations for each box size, and then matching them to produce a
smoothed power spectrum. This reference $P(k)$ can be used as one test
of the emulator's accuracy.

\begin{figure}[t]
\begin{center}
\includegraphics[width=80mm]{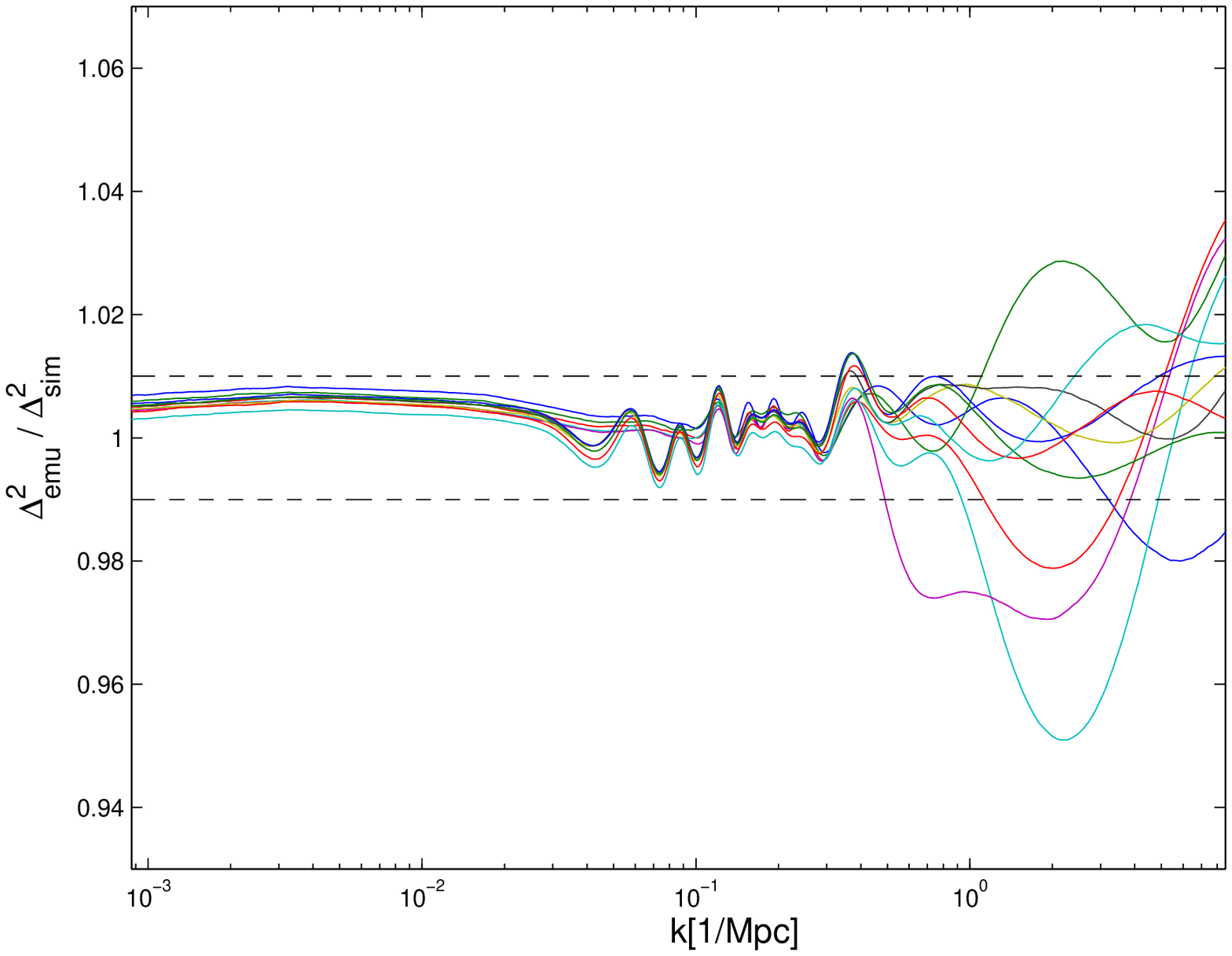}
\includegraphics[width=80mm]{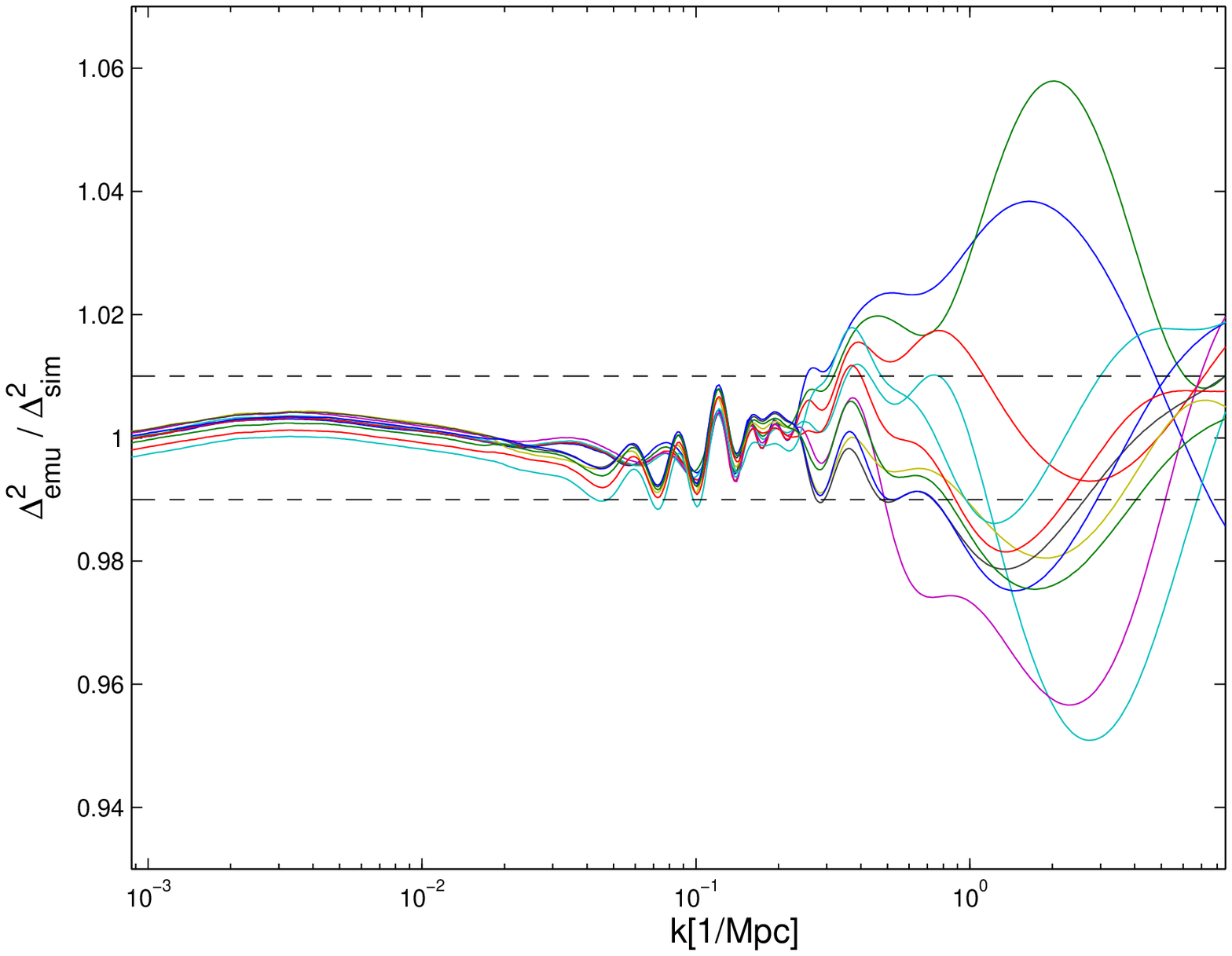}
\end{center}
\caption{Accuracy of the emulator predictions for the reference
  $\Lambda$CDM model, M000, not used to construct the emulator. The
  ratio of the emulator prediction to the smoothed M000 simulation
  result is shown as a function of $k$. In the upper panel we show the
  predictions for the emulator with the Hubble parameter $h$ fixed to
  the best-fit CMB value for each model, in the lower panel, the
  results when $h$ is allowed to be an independent parameter.  The
  different colors represent results at different redshifts: blue:
  $a=0.2$, green: $a=0.25$, red: $a=0.2857$, cyan: $a=0.3333$, purple:
  $a=0.4$, yellow: $a=0.5$, gray: $a=0.6$, blue: $a=0.7$, green:
  $a=0.8$, red: $a=0.9$, cyan: $a=1.0$. The sequence of $a$ values
  correspond to the redshifts, $z=4$, 3, 2.5, 2, 1.5, 1, 0.67, 0.43,
  0.25, 0.11, 0.0. The dashed horizontal line indicates the $1\%$
  accuracy limit. Allowing for $h$ to vary freely only mildly affects
  the emulator error behavior. }
\label{m000}
\end{figure}

Figure~\ref{m000} shows results for two emulator versions, one in
which the value of $h$ is locked by the CMB distance to last
scattering constraint for each model, as in the original
emulator \citep{coyote3} (upper panel), and a second version where $h$
is allowed to be a free parameter following the approach presented in
Section~\ref{sec:hubble} (lower panel). Both panels show the ratio of
the emulator to the simulation result as a function of $k$, the
different solid curves correspond to eleven different redshifts
between $z=4$ and $z=0$ used to build the emulator. In the $h$-fixed
case, over a wide $k$ range, and for all redshifts, the emulator
prediction is accurate at the 1\% level, degrading slightly only at
larger $k$ values, but still remaining better than 5\%.

When $h$ is allowed to range freely (lower panel), then over the $k$
range where RPT is used (up to $k\sim0.03~$Mpc$^{-1}$ for the low
redshift results), the result is excellent, better than that in the
top panel. This is not surprising since for this range we now have 100
models to predict the power spectrum. Beyond that matching point, the
accuracy degrades slightly in the quasi-linear regime compared to the
more restricted emulator, but is still around 1\%. Finally, beyond
$k\sim1~$Mpc$^{-1}$, the predictions are less accurate, but the worst
case error remains around 5\%.

\begin{figure}[t]
\begin{center}
\includegraphics[width=80mm]{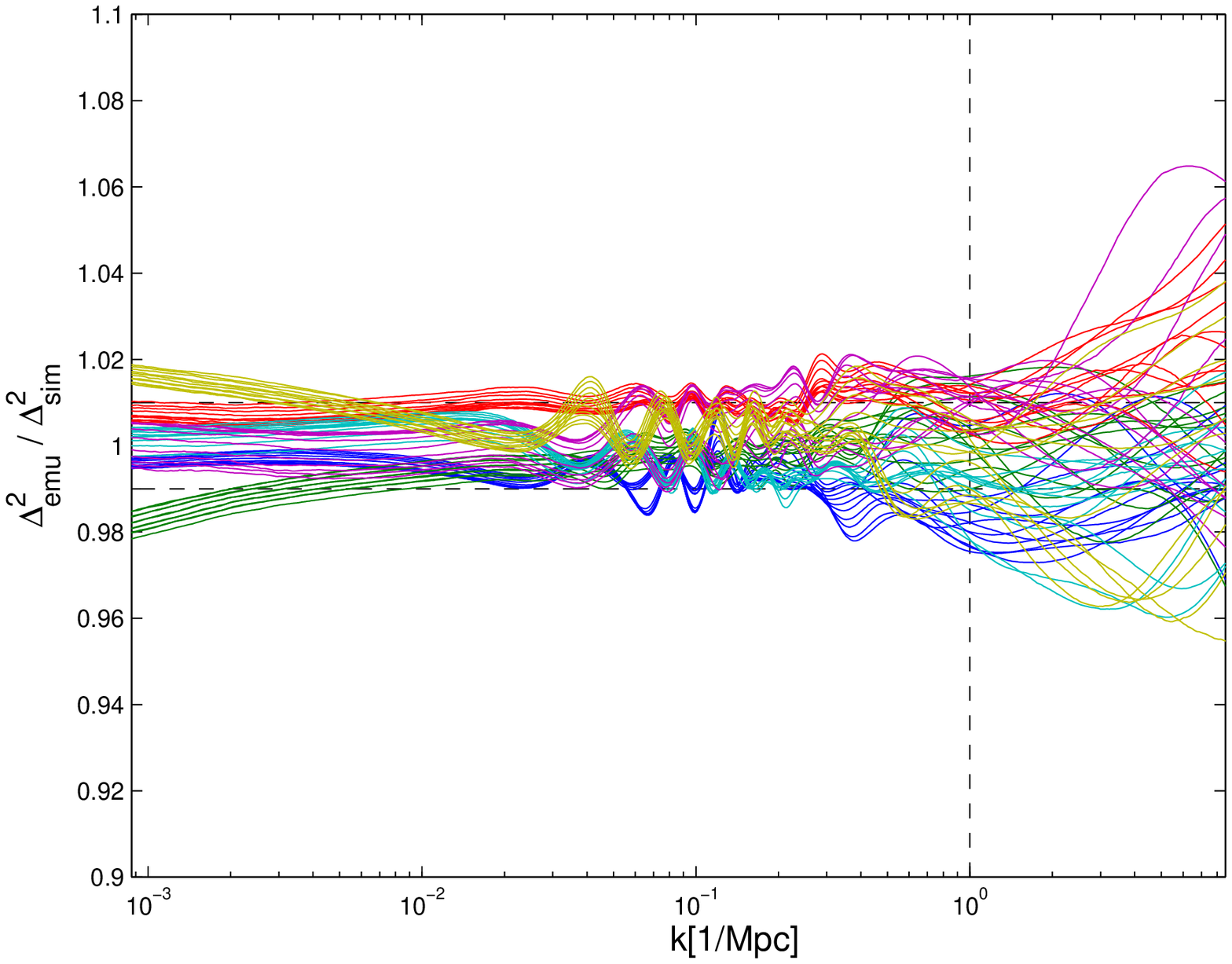}
\includegraphics[width=80mm]{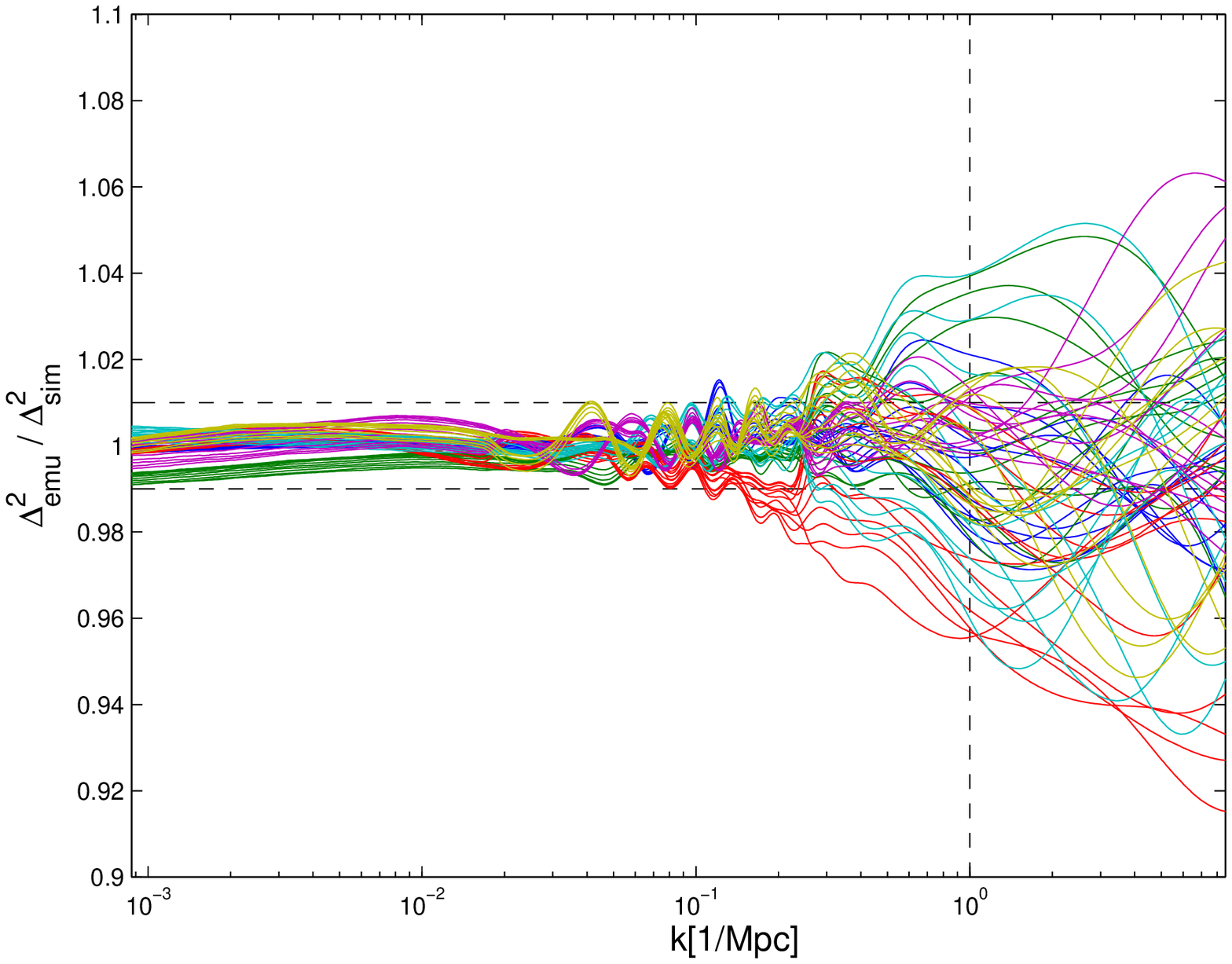}
\end{center}
\caption{Holdout test for six models. For every test, the chosen model
  is excluded while building the emulator. The emulator prediction is
  then compared with that of the `held out' model. The ratio of the
  emulator prediction to the smoothed simulation result for the six
  reference models is shown. As in Fig.~\ref{m000}, the top panel
  shows results with $h$-fixed emulators, while in the bottom panel,
  this constraint is relaxed. The different colors show results for
  different models (for each model we show results at all redshifts):
  blue: M004, green: M008, red: M013, cyan: M016, purple: M020,
  yellow: M026. The dashed horizontal line indicates the 1\% accuracy
  limit, while the dashed vertical line shows $k=1$~Mpc$^{-1}$, the
  limit of our previous emulator \citep{coyote3}.}
\label{holdouts}
\end{figure}

Next, we present results from a number of `holdout' tests. In these
tests, one model out of the 37 is excluded and an emulator is
constructed based on the remaining 36 models. Then with this new
emulator, a prediction for the excluded model is generated and the
accuracy of the prediction determined. This test has an obvious
caveat, particularly relevant in cases where not many models are
available -- degradation of the emulator quality because an important
point in the parameter-space hypercube is omitted. This can be
particularly serious if the excluded point is on the edge of the
design, because now extrapolation to the edge of the hypercube is
required, which has a much higher error. In order to avoid this
additional complication, we restrict our holdout test to models that
lie well within the hypercube. Nevertheless, it should be kept in mind
that reported errors in the holdout tests can be larger than the
actual values. The holdout test results, shown in Fig.~\ref{holdouts}
are similar to the ones found for M000, both in error amplitude and
trends with increasing $k$. In the case where $h$ is allowed to be a
free variable, the predictions are again less accurate, already in the
quasi-linear regime, and degrade further beyond $k\sim1~$Mpc$^{-1}$,
to of order 5\%. Note that the low-$k$ error band is consistent with
that for the linear theory test case (Cf. Fig.~\ref{hubble_test}).

To summarize, the accuracy of the new, extended, emulator is
well-characterized at all redshifts. With $h$ fixed to the CMB
constraint, the accuracy to $k\sim 1~$Mpc$^{-1}$ is at the $\sim1\%$
level, degrading to $\sim 5\%$ for $k\sim 10~$Mpc$^{-1}$. When $h$ is
allowed to be a free parameter, the degradation in accuracy is
relatively modest.

In order to improve the accuracy of the emulator with $h$ included as
a free parameter, a new design with more data points would have to be
created.  Based on the convergence tests in \cite{HHHNW}, where the
design space was varied between 32 and 128 models, a design with
around 40 to 50 cosmological models should lead to percent level
errors. We did not follow this strategy in the current paper because
it would not allow us to use the simulation data already obtained in
our previous work. Some of these questions (e.g., how to create nested
designs that allow improvements of the emulator accuracy by adding
more models in a systematic fashion) are currently under
investigation, and will be addressed in future work.

\subsection{Sensitivity Investigation}

\begin{figure}[t]
\begin{center}
\includegraphics[width=90mm]{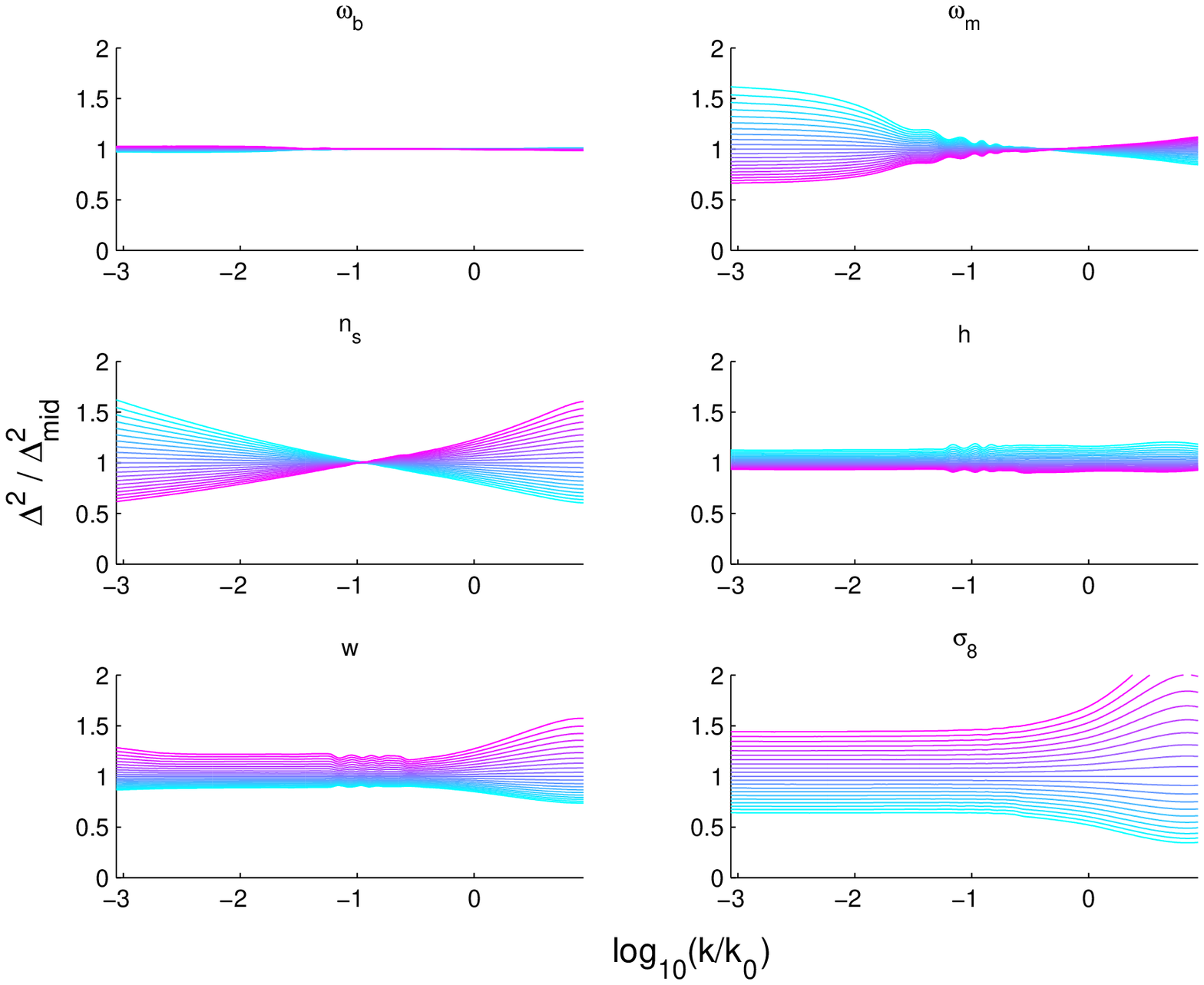}
\includegraphics[width=90mm]{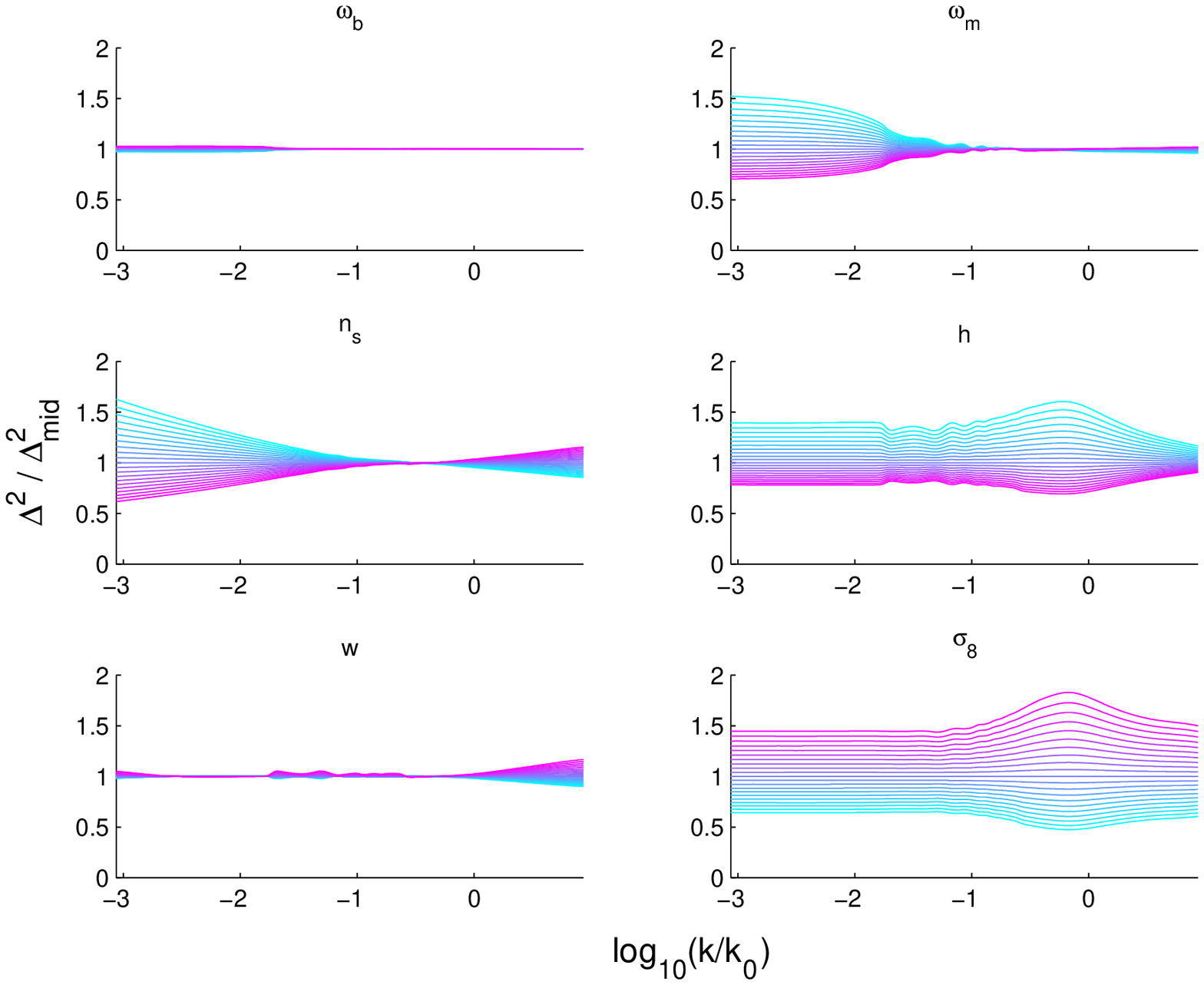}
\end{center}
\caption{Same as in Fig.~\ref{sens} but for the nonlinear power
  spectrum. The upper panel shows results at $z=4$ while the lower
  panel shows results at $z=0$.}
\label{sensnl}
\end{figure}

With the full emulator at hand we can now carry out a sensitivity
analysis for the nonlinear power spectrum, along the same lines as
conducted in Section~\ref{sec:hubble} for the linear power
spectrum. As before, we fix five of the six cosmological parameters at
the midpoints of their range and then vary the sixth parameter between
its minimum and maximum value. The results -- for the lowest and
highest redshift we consider -- are shown in Fig.~\ref{sensnl}. The
upper six panels show the results at $z=4$ and the lower six panels at
$z=0$.

Some interesting features in the results can be noted. As to be
expected, the baryon fraction $\omega_b$ does not affect the power
spectrum significantly. The best constraints we have for $\omega_b$
come from CMB measurements; Planck determines this parameter to
exquisite accuracy. The effect of the matter fraction $\omega_m$
dominates at large scales, here again, CMB results deliver very good
estimates of this parameter. In general, it is important to allow for
variations in $\omega_m$ due to degeneracy issues; $\omega_m$
influences the overall amplitude of the power spectrum as do
$\sigma_8$ and $h$. The spectral index $n_s$, influencing the tilt of
the power spectrum also mainly effects large scale behavior. The
remaining three parameters, $h$, $w$, and in particular $\sigma_8$,
alter the power spectrum on all scales and constraints on these
parameters can therefore be improved by measurements of nonlinear
scales.

In particular, the influence of $\sigma_8$ at the two redshifts shown
in Fig.~\ref{sensnl} on the power spectrum is noteworthy.  At high
redshift, nonlinearities for models with large $\sigma_8$ (pink) have
already developed considerably, therefore the ratio with the midpoint
power spectrum shows large values. The power spectra for small values
of $\sigma_8$ (blue) show only mild nonlinear growth and therefore the
ratio is still almost flat. Over time, the nonlinear turn-over moves
in to smaller $k$ ranges (affecting larger and larger scales), causing
the bump just before $k=1$~Mpc$^{-1}$ shown in the panel for $z=0$.
At this epoch the overall difference between the models has decreased,
since at this point all models have reached the nonlinear regime. A
similar observation can be made for $w$ -- at early times the
differences between the models are more pronounced than at later
times. It is easy to imagine using the new emulator as a convenient
tool to investigate the effects of varying cosmological parameters and
redshift on $P(k)$ in the nonlinear regime.

\subsection{Comparison with Halofit}
\label{halofit}

\begin{figure}[t]
\begin{center}
\includegraphics[width=80mm]{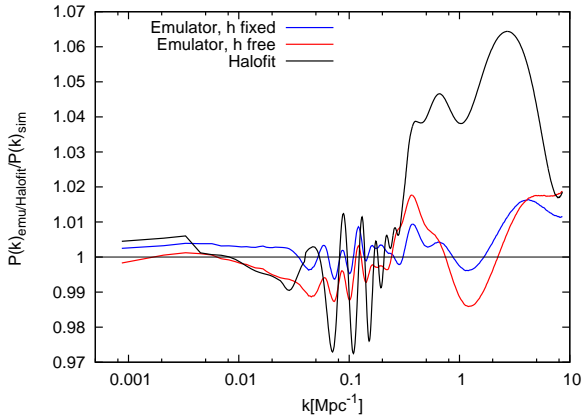}
\caption{Performance of the emulator in comparison with Halofit
  from \cite{takahashi} for M000 at $z=0$. Shown are the ratios of the
  emulator with fixed Hubble parameter and the smoothed simulation
  (blue), with the free Hubble parameter emulator (red), and with
  Halofit (black). (The emulator results are shown also in
  Fig.~\ref{m000} in cyan and are here repeated for easy comparison
  with Halofit). The $h$-fixed emulator is accurate at the percent
  level out to $k\sim 1~$Mpc$^{-1}$ and at the 2\% level at higher $k$, the
  $h$-free emulator is accurate at the 2\% level throughout. Halofit
  on the other hand shows deviations at the 3\% level at
  $k\sim0.1~$Mpc$^{-1}$ and up to 6\% in the higher $k$ regime.}
\label{fig:m000_comp}
\end{center}
\end{figure}

\begin{figure}[t]
\begin{center}
\includegraphics[width=60mm,angle=270]{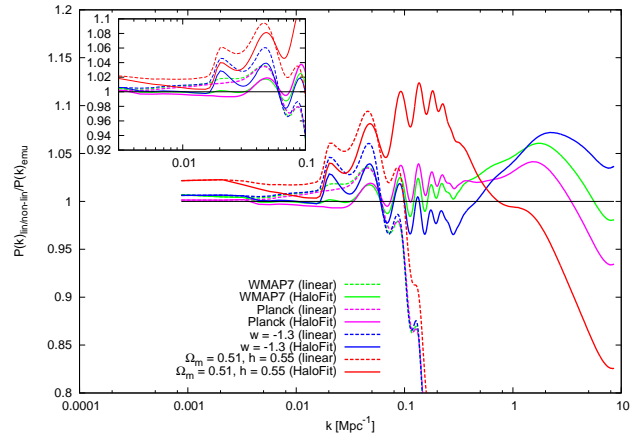}
\caption{Performance of the emulator in comparison with linear theory
  (dotted) supplied by CAMB and Halofit from \cite{takahashi}
  (solid). Unless otherwise stated in the legend, the cosmological
  parameters are: $\omega_m = 0.1296, \omega_b= 0.0224, n_s = 0.97,
  \sigma_8 = 0.8, w = -1, h = 0.72$. The parameters for WMAP-7 and
  Planck are given in Eq.~(\ref{eq:params}).  See text for
  discussion.}
\label{fig:halofit}
\end{center}
\end{figure}

We now perform a comparison between the emulator and Halofit, a
popular fitting formula for the matter power spectrum motivated by the
halo model. We use the updated version with 35 fitting parameters
provided by \cite{takahashi}, which is calibrated to larger volume and
higher resolution N-body simulations than the original \cite{Smith03}
formula. It is based on simulations run on six WMAP cosmologies (years
1, 3, 5, 7, and two WMAP7 models, but with $w=-0.8$ and $w=-1.2$, instead
of $w=-1$). The obtained fitting formula was checked against
simulations run using the first 10 of the 37 Coyote models. This
improved version of Halofit is stated to be accurate to $\sim 5\%$ in
the range $0 < k < 1$~$h$Mpc$^{-1}$ and $\sim 10\%$ at $1 < k <
10$~$h$Mpc$^{-1}$, for $0<z<3$.

One important feature of the emulator-based approach is that, close to
the parameter values of the underlying simulations, the emulator error
is small, if not essentially the simulation error. However, this is
not the case with a fitting formula. For instance, even though the
N-body simulations that underlie the new Halofit formula agree with
the original Coyote emulator at the $\sim1-2\%$ level over its range
of validity for the 10 Coyote models used as a reference
in \cite{takahashi} (se Fig.~2 of the cited reference), this good
level of agreement drops to 8-13\% (as a function of the $k$ range)
when the fitting formula is used in the comparison. Depending on what
one may be trying to achieve, this loss of accuracy in Halofit is
significant.

We first compare the predictions of the emulator (both versions, $h$
free and $h$ fixed) at $z=0$ for M000 against the smooth simulations
(in the same way as shown in Fig.~\ref{m000}) and the prediction of
Halofit for the same model. Since the emulator was built without
including this simulation, this is a good test of both prediction
schemes, emulator and Halofit. The results are shown in
Fig.~\ref{fig:m000_comp}. The emulator predictions for the $h$-fixed
version are accurate at 1\% out to $k\sim 1~$Mpc$^{-1}$ and then degrade very
slightly to 2\%.  The $h$-free version of the emulator is accurate at
the 1-2\% level throughout.  Halofit shows 3\% deviations compared to
the simulation in the mildly nonlinear regime ($k\sim 0.1~$Mpc$^{-1}$) and
more than 6\% in the nonlinear regime.

Next we show a comparison of the emulator ($h$-free version only)
directly with Halofit for a variety of cosmologies at $z=0$.
Figure~\ref{fig:halofit} displays the ratio of the emulator with
respect to linear theory power spectra from CAMB and Halofit as
modified by \cite{takahashi}. We choose to test a WMAP7 \citep{wmap7}
and a Planck \citep{planck} cosmology and two other cosmologies on the
limits of the design of our emulator. The red and blue curves in
particular are on the edge of the new design with $h$ as a free
parameter. The cosmological parameters for these two model are the
same as model M000: $\omega_m = 0.1296, \omega_b = 0.0224, n_s = 0.97,
\sigma_8 = 0.8, w = -1, h = 0.72$ and in our comparison we vary one to
two parameters at a time as noted on the legend.

At large scales, $k < 0.01$~$h$Mpc$^{-1}$, all power spectra are
consistent at the percent level; this is less trivial than it seems
since Halofit and linear theory start diverging at $k\sim
0.003$~$h$Mpc$^{-1}$ and we use the full version of the emulator with
extreme values of $h$ located at the edge of the design.  On
quasi-linear scales, there is $\sim 5$\% deviation with some
oscillatory structure from the BAO feature.  On smaller scales, the
accuracy is consistent with what is stated in \cite{takahashi}, apart
from the matter dominated case with $\Omega_m$ = 0.51, in which the
Halofit power spectrum is suppressed by more than 20\%. The relatively
uniform error behavior of an emulator -- a result of the sampling
theory and the Gaussian process based-regression methodology -- is
hard to reproduce in a fitting formula; the last result of the
comparison provides an example of this.

\section{Conclusion and Outlook}
\label{sec:conc}

High accuracy predictions for cosmological observables in the
nonlinear regime will be crucial in the future to exploit the power of
ongoing and upcoming cosmological surveys. These predictions have to
at least match, better yet, exceed the accuracy of the measurements
themselves. Achieving this goal with first principles predictions or
perturbation theory is impossible because of the highly nonlinear and
dynamical nature of the problem. High-accuracy predictions must
therefore be obtained from state of the art simulations.

In a series of recent papers (the Coyote Universe papers) we have
shown how to build a prediction tool from a relatively small set of
simulations (37 cosmological models) to generate the matter power
spectrum out to $k\sim 1$~Mpc$^{-1}$ at the 1\% accuracy out to
$z=1$. In the current paper, using what we consider the `minimal'
amount of work, we have extended the emulator to significantly higher
values of redshift and wavenumber. In addition, we have added one new
free parameter, $h$.

With current computational resources, it is impossible to obtain high
emulation accuracy out to large $k$ with a collection of single-box
simulations, because the dynamic range requirements impose a very high
computational cost. The use of nested boxes avoids the computational
cost, at the expense of reduced accuracy, resulting from increased
sampling variance as well as reduced accuracy in the region of the
nonlinear turnover. Nevertheless, with this approach we have improved
on the accuracy attained by any other existing prediction
tool. Overall, our predictions for the power spectrum are accurate to
better than 5\%, and over smaller $k$ ranges around 1\%, for
$0<z<4$. In the high-$k$ regime, baryonic effects will have to be
included in any case, so any future strategy should include methods
for improving the accuracy of the gravity-only simulation results, as
well as ways to model the baryonic contributions.

In a companion paper we will present a method for extrapolating the
results beyond $k\sim 10$~Mpc$^{-1}$ (including estimates of the
associated errors) and an emulator for the shear power spectrum and
other weak lensing observables. More detailed work relevant for
surveys (DES, LSST) is also underway.

For future surveys, the cosmological parameter space will have to be
enlarged beyond what was considered here. In the case of DES, for
example, predictions for dynamical dark energy models are
needed. Increasing the sample space will further increase the
associated computational costs. On the positive side, results from
surveys such as Planck help to narrow down the parameter ranges
substantially, which in turn will help to reduce the number of models
we have to investigate. Multi-level sampling schemes can be devised to
deal with these situations, one of our directions for future work.

\appendix

\section{Appendix A: Test of the New Emulator Against Large
  High-resolution HACC Simulation} 
\label{appendixa}

In this Appendix we present an additional test of our new $h$-free
emulator set-up by using a simulation that covers a large range of the
$k$ values of interest and uses a different N-body code, namely HACC
(Hardware/Hybrid Accelerated Cosmology Code)~\citep{habib12}. The
cosmology for this test is the best-fit WMAP-7 cosmology
($\omega_m=0.1335,~\omega_b=0.02258,~n_s=0.963~,
w=-1.0,~\sigma_8=0.8~,h=0.71$).  With this final test we demonstrate
that (i) the matching strategy described in Section~\ref{sec:nested}
works well at the accuracy reported in the paper, (ii) the smoothing
procedure described in Section~\ref{sec:smooth} does not introduce any
biases, and that (iii) the power spectrum calculation is robust under
different N-body implementations. The last point had already been made
in the Coyote-I paper~(\citealt{coyote1}) by comparing {\sc Gadget-2}
results with the ART code (see Fig.~4 in \citealt{coyote1}), and is
here again confirmed with a third code out to higher $k$ values.

\begin{figure}[t]
\begin{center}
\includegraphics[width=90mm]{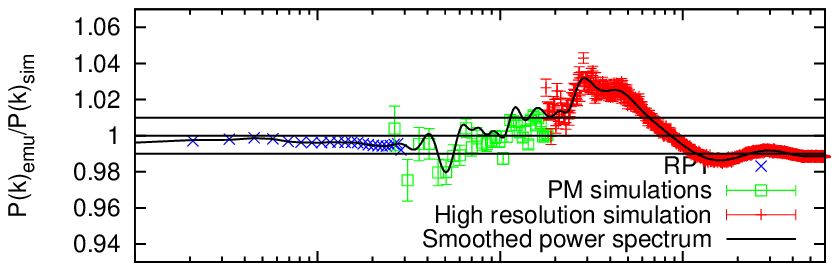}
\includegraphics[width=90mm]{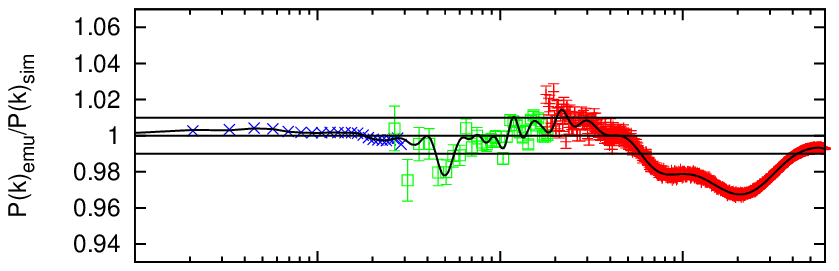}
\includegraphics[width=90mm]{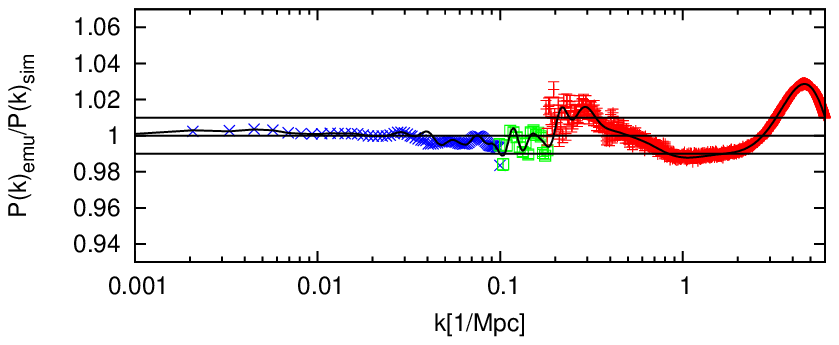}
\caption{Comparison of the $h$-free emulator with raw simulations,
  i.e., with no smoothing applied at three redshifts (upper panel:
  $z=0$, middle panel: $z=1$, lower panel: $z=2$). We show the results
  from resummed perturbation theory in blue, the average of 16 PM runs
  in green, and the result from one high resolution simulation in
  red. The black lines show the result from matching and smoothing all three pieces
  following the procedure described in detail in Section~\ref{sec:smooth}. 
  Error bars describe the error in the simulations due to the
  finite number of modes, or cosmic variance.  The predictions are
  well within the expected accuracy of the emulator.\label{simcomp} }
\label{fig:rat}
\end{center}
\end{figure}

We show results for three redshifts, $z=0,1,2$ -- at higher redshifts
the large simulation volume leads to excess particle shot noise in the
higher $k$ regime which is of most interest for this test. We cover a
$k$ range between 0.001 to 6~Mpc$^{-1}$ via (i) renormalized
perturbation theory out to $k\sim0.04$~Mpc$^{-1}$ for $z=0,~1$ and
$k\sim0.14$~Mpc$^{-1}$ for $z=2$ as in the main paper; (ii) 16
realizations of PM simulations with 512$^3$ particles on a 1024$^3$
grid in a (1300~Mpc)$^3$ volume out to $k\sim0.25$~Mpc$^{-1}$ also as
in the main paper; (iii) a (2100Mpc)$^3$ volume, high-resolution
simulation with 3200$^3$ particles and 6.5~kpc force resolution to
cover the remaining $k$-range. The difference with the main paper
arises in the intermediate regime, where we drop the second set of PM
simulations since the high-resolution simulation covers a
significantly larger volume than the {\sc Gadget-2} runs. The
realization scatter for the smaller $k$ values is reduced to a
low-enough level that the single simulation spans the quasi-linear
regime.  Figure~\ref{simcomp} shows the comparison of the emulator
with the unsmoothed and smoothed simulations.  This figure should be compared to
Fig.~\ref{m000}, lower panel, where we show a similar cosmology (M000)
but using the matching strategy of different size simulations that has
been employed throughout the paper. The error level is very similar in
both cases, demonstrating very good performance of the matching and
smoothing strategy, as well as excellent agreement between HACC and
{\sc Gadget-2}.

\section{Appendix B: Improvement of Original Emulator}
\label{appendixb}

In the course of the work presented in this paper, some changes were
made to previous versions of the CosmicEmu from  \cite{coyote3}.

First,  \cite{Taruya} present work on perturbation theory and compare
their results to the CosmicEmu for the cosmologies that were simulated
to build the CosmicEmu.  Overall, they note a close match, but for
M015, they find a notable difference.  In the course of investigating
this issue, we discovered a typographical error in the input files
that were used to build the CosmicEmu.  The file containing the input
parameters differed from the input parameters used to run the actual
simulations for three cosmologies: M015, M017, and M019.  The effects
of this error were quite local to the neighborhood of these three
cosmologies.

Second, we experimented with the prior parameters for the Gaussian
process emulation procedure.  In order to explain this more fully, we
must describe one more subtle detail in Gaussian process fitting.
Appendix B in \cite{coyote2} describes the statistical model for the
Gaussian process emulation.  Equation B1 in \cite{coyote2} gives the
Gaussian process distribution for the principal component weights with
covariance matrix $\Sigma = \lambda_{w_i}^{-1} R(\theta,
\theta^{\prime}; \rho_{w_i})$.  The function $R(\cdot)$, given in
Equation B2, results in a response that interpolates and is extremely
smooth.  As a result, the model can be susceptible to estimation
problems and overfitting.  Thus, in practice, we include an additional
term in the covariance specification to relax the interpolation
requirement slightly.  The actual covariance is given by $\Sigma =
\lambda_{w_i}^{-1} R(\theta, \theta^{\prime}; \rho_{w_i}) +
\lambda_{nug}^{-1} I$, where $I$ is the identity matrix.  The new
parameter $\lambda_{nug}$ is an additional precision term (the
`nugget') that governs how well the estimated response function
interpolates the data.  When this parameter is large, the response
surface interpolates better.  We estimate this parameter along with
all of the others, so the data can inform about the smoothness of the
model.  Like the other precision parameters described in
\cite{coyote2}, it has a Gamma prior distribution (see Equation B5 for
an example).  Unfortunately, the default prior parameters in the
estimation software can prevent these precisions from being as large
as the data would allow and prevent the response surface from
capturing all of the behavior of the simulations.  We now set these
parameters at $a_{nug,i}=1$ and $b_{nug,i}=0$ where $i$ indexes the
principal components.  This prior gives the data much more control
over the parameter estimate.  The new estimates for $\lambda_{nug}$
are significantly larger than the previous estimates.  The resulting
response surface for each principal component basis is more accurate
and more principal component basis functions can be used (seven
instead of five).

\begin{figure}[t]
\begin{center}
\includegraphics[width=90mm]{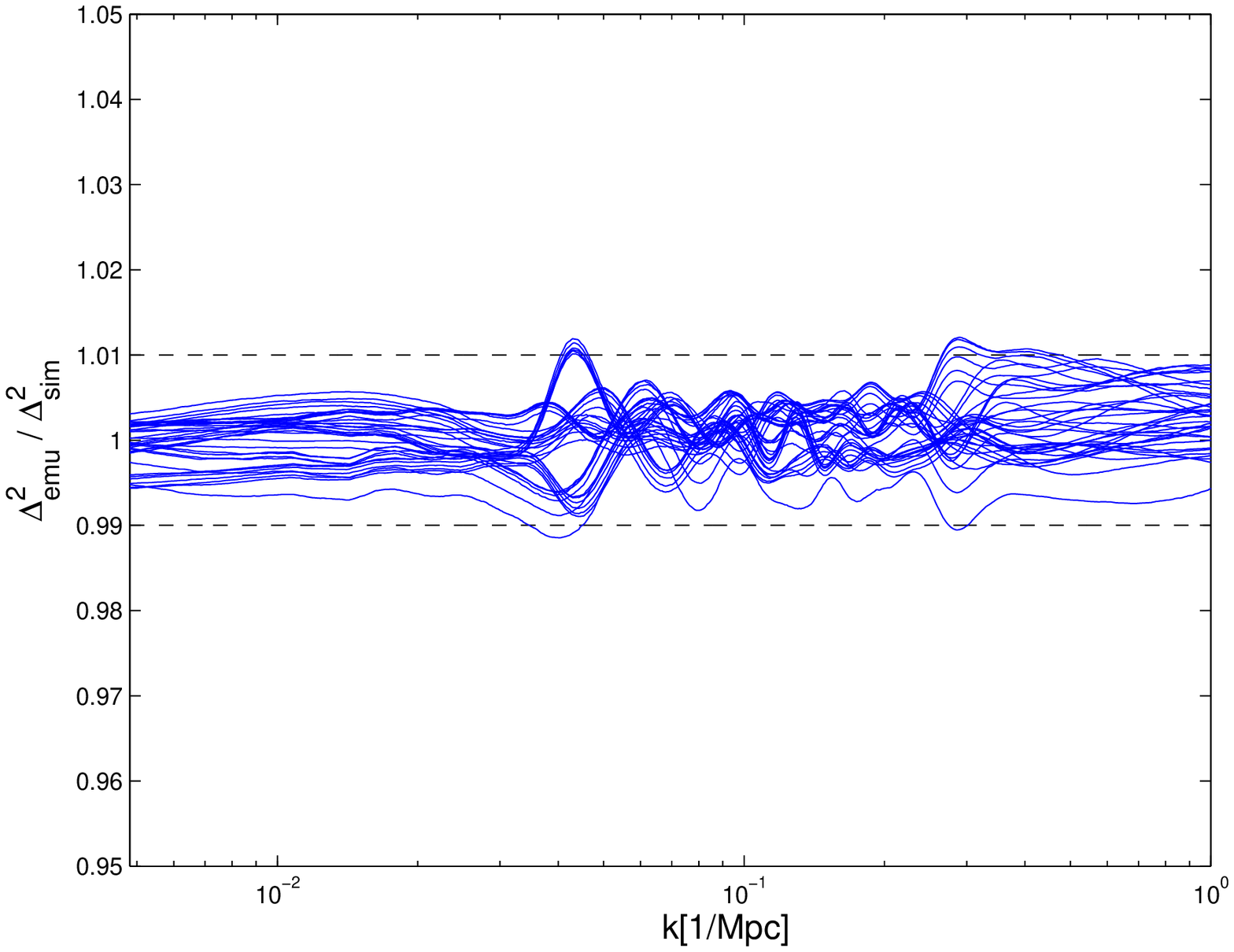}
\includegraphics[width=90mm]{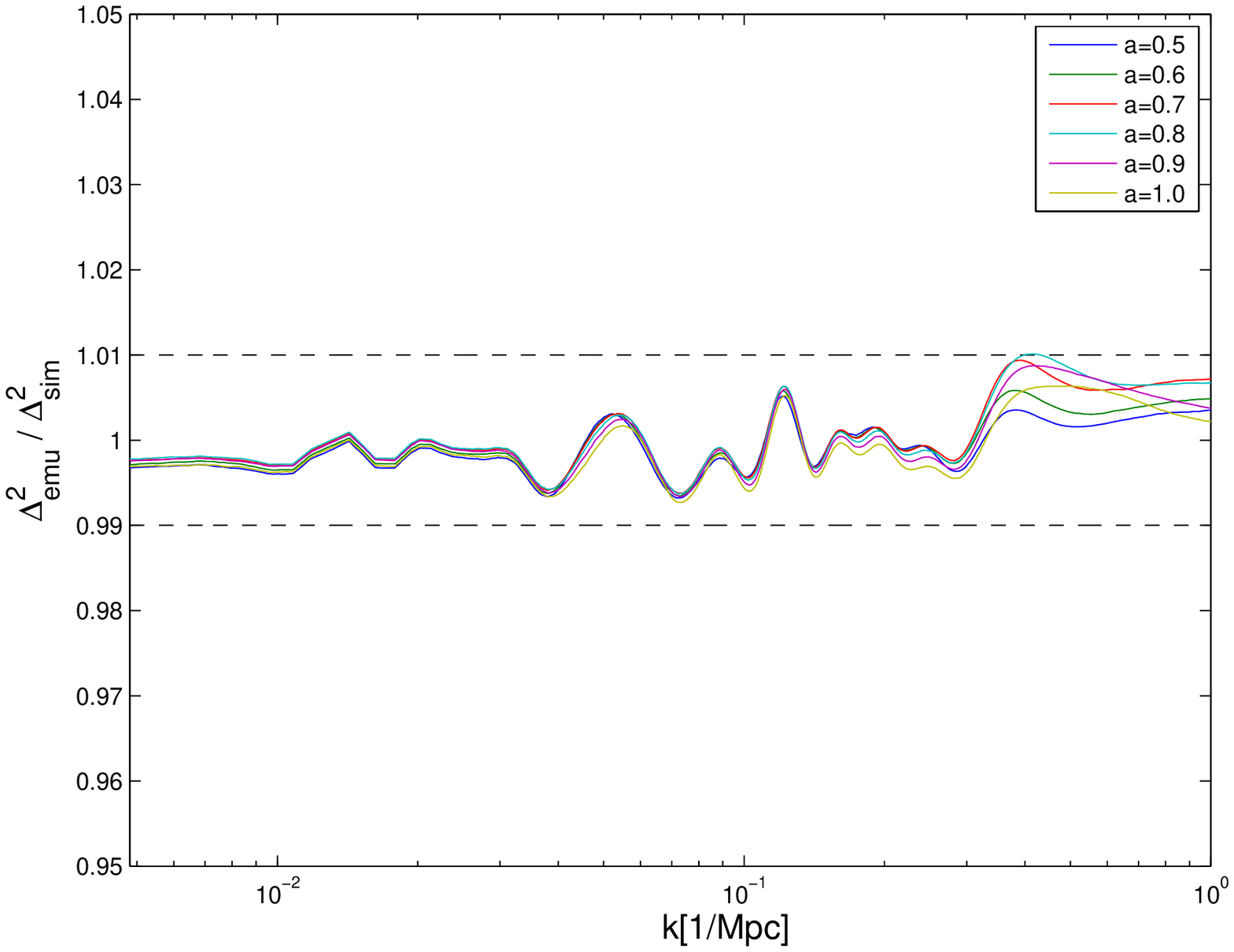}
\end{center}
\caption{Holdout (top) and M000 (bottom) results for the improved fit
  to the original CosmicEmu described in \cite{coyote3}.  Compare to
  Figures 9 and 10 in that paper.}
\label{coyotefixed}
\end{figure}

Figure \ref{coyotefixed} shows the results of these two improvements
on the holdout predictions and the prediction for the best fit
cosmology M000.  The fits, while good before, are now further
improved.

\begin{acknowledgments}

  Part of this research was supported by the DOE under contract
  W-7405-ENG-36. Partial support for this work was provided by the
  Scientific Discovery through Advanced Computing (SciDAC) program
  funded by the U.S. Department of Energy, Office of Science, High
  Energy Physics, and Advanced Scientific Computing Research. We are
  grateful to Martin White and Christian Wagner for their important
  contributions to the original Coyote Universe series which forms the
  basis of this new work. We thank Tim Eifler and Adrian Pope for
  useful discussions. We would like to thank Volker Springel for
  making {\sc Gadget-2} publicly available and Jordan Carlson for the
  Copter code.  We would also like to thank Taruya et al. for making
  their perturbation code publicly available, as it was very helpful
  in double-checking some of our results.  
  
  We are grateful for
  computing time granted to us as part of the Los Alamos Open
  Supercomputing Initiative.  This research used resources at NERSC
  (National Energy Research Scientific Computing Center), which is
  supported by the Office of Science of the U.S. Department of Energy
  under Contract No. DE-AC02-05CH11231. This research used resources 
  of the ALCF, which is supported by DOE/SC
under contract DE-AC02-06CH11357

The submitted manuscript has been created by UChicago Argonne, LLC,
Operator of Argonne National Laboratory (``Argonne''). Argonne, a
U.S. Department of Energy Office of Science laboratory, is operated
under Contract No. DE-AC02- 06CH11357. The U.S. Government retains for
itself, and others acting on its behalf, a paid-up nonexclusive,
irrevocable worldwide license in said article to reproduce, prepare
derivative works, distribute copies to the public, and perform
publicly and display publicly, by or on behalf of the Government.

\end{acknowledgments}


\begin{thebibliography}{99}

\bibitem[{{Abell et al.}(2009)}]{lsst}
Abell,~P.A. et al. [LSST Science Collaborations and LSST Project
Collaboration], arXiv:0912.0201 [astro-ph.IM]

\bibitem[{{Ade et al.}(2013)}]{planck}
Ade,~P.A.R. et al. [ Planck Collaboration], arXiv:1303.5076
[astro-ph.CO] 

\bibitem[{{Agarwal et al.}(2012)}]{feldman} 
Agarwal,~S., Abdalla,~F., Feldman,~H., Lahav,~O., \&
Thomas,~S.A. 2012, MNRAS, 424, 1409 

\bibitem[{{Baugh, Gaztanaga, \& Efstathiou}(1995)}]{baugh95}
Baugh,~C.M., Gaztanaga,~E., \& Efstathiou,~G. 1995, MNRAS, 274, 1049  

\bibitem[{{Bhattacharya et al.} (2011)}]{bhattacharya11}
Bhattacharya,~S., Heitmann,~K., White,~M., Luki\'c,~Z., Wagner,~C., \&
Habib,~S. 2011, \apj, 732, 211  

\bibitem[{{Bird, Viel, \& Haehnelt}(2012)}]{bird} 
Bird,~S., Viel,~M., \& Haehnelt,~M.G. 2012, MNRAS, 420, 2551 

\bibitem[{{Carlson, White, \& Padmanabhan}(2009)}]{copter}
Carlson,~J., White,~M., \& Padmanabhan,~N. 2009, Phys. Rev. D80,
043531

\bibitem[{{Cunha \& Evrard}(2010)}]{cunha10} 
Cunha,~C.E. \& Evrard,~A.E. 2010, Phys. Rev. D., 81, 083509 

\bibitem[{{Crocce \& Scoccimarro}(2006)}]{crocce1}
Crocce, M. \& Scoccimarro, R. 2006, Phys. Rev. D73, 063519

\bibitem[{{Crocce \& Scoccimarro}(2008)}]{crocce2}
Crocce, M. \& Scoccimarro, R. 2008, Phys. Rev. D77, 023533

\bibitem[{{Das et al.}(2012)}]{act} 
Das,~S. et al. 2011, Phys. Rev. Lett., 107, 021301

\bibitem[{{de Jong et al.}(2013)}]{kids}
de Jong, J.T.A, et al. 2013, Experimental Astronomy, 35, 25

\bibitem[{{Drinkwater et al.}(2010)}]{wigglez} 
Drinkwater,~M.J. et al. 2010, MNRAS, 401, 1429

\bibitem[{Gnedin, Kravtsov, \& Rudd}(2011)]{gnedin11} 
Gnedin,~N.Y., Kravtzov,~A.V., \& Rudd,~D.H. 2011, ApJS, 194, 46

\bibitem[{{Eifler}(2011)}]{eifler} 
Eifler,~T. 2011, MNRAS, 418, 536

\bibitem[{{Habib et al.}(2007)}]{HHHNW} 
Habib,~S., Heitmann,~K., Higdon,~D., Nakhleh,~C., \&
Williams,~B. 2007, Phys. Rev. D, 76, 083503

\bibitem[{{Habib et al.}(2012)}]{habib12} 
Habib,~S., Morozov,~V., Finkel,~H., Pope,~A., Heitmann,~K.,
Kumaran,~K., Peterka,~T., Insley,~J., Daniel,~D., Fasel,~P.,
Frontiere,~N., \& Luki\'c,~Z. 2012, Proc. SC12, arXiv:1211.4864

\bibitem[{{Hearin \& Zentner}(2011)}]{hearin} 
Hearin,~A., Zentner,~A.R., \& Ma,~Z. 2012, JCAP, 1204, 034 

\bibitem[{{Heitmann et al.}(2006)}]{HHHN} 
Heitmann,~K., Higdon,~D., Nakhleh,~C., \& Habib,~S. 2006, ApJ, 646, L1 

\bibitem[{{Heitmann et al.}(2009)}]{coyote2} 
Heitmann~K., Higdon~D., White~M., Habib~S., Williams,~B.J., \&
Wagner,~C. 2009, ApJ, 705, 156

\bibitem[{{Heitmann et al.}(2010)}]{coyote1}
Heitmann~K., White~M., Wagner~C., Habib~S., \& Higdon~D. 2010, ApJ,
715, 104 

\bibitem[{{Higdon}(2002)}]{Higdon2002}
Higdon,~D. 2002, in Quantitative Methods for Current Environmental
Issues, Space and Space-Time Modeling using Process Convolutions,
ed. C.~Anderson, V.~Barnet, R.C.~Chatwin, \& A.H.~El-Shaarawi (Berlin:
Springer), 37 

\bibitem[{{Huff et al.}(2011)}]{huff} 
Huff,~E.M., Eifler,~T., Hirata,~M., Mandelbaum,~R., Schlegel,~D., \&
Seljak,~U. 2011, arXiv:1112.3143 [astro-ph.CO]

\bibitem[{{Huterer \& Takada}(2005)}]{huttak} 
Huterer,~D. \& Takada,~M. 2005, Astropart. Phys. 23, 369 

\bibitem[{{Jee et al.}(2013)}]{dls}
Jee, M. J., Tyson, J. A., Schneider, M.D., Wittman, D., Schmidt, S.,
\& Hilbert, S. 2013, ApJ, 765, 26

\bibitem[{{Jenkins et al.}(2001)}]{jenkins01}
 Jenkins,~A. et al. 2001, MNRAS 321, 372

\bibitem[{{Jing et al.}(2006)}]{jing06}  
Jing,~Y.P., Zhang,~P., Lin~W.P., Gao~L., \& Springel~V. 2006, ApJ,
640, L119 

\bibitem[{{Kilbinger et al.}(2012)}]{cfhtlens}
Kilbinger,~M. et al. 2012, arXiv:1212.3338 [astro-ph.CO] 

\bibitem[{{Komatsu et al.}(2011)}]{wmap7} 
Komatsu,~E. et al. 2011, ApJS, 192, 18

\bibitem[{{Kwan et al.}(2013)}]{kwan12} 
Kwan,~J., Bhattacharya,~S., Heitmann,~K., \& Habib,~S. 2013, ApJ, 768,
123 

\bibitem[{{Lawrence et al.}(2010)}]{coyote3} 
Lawrence,~E., Heitmann,~K., White~M., Higdon~D., Wagner~C., Habib~S.,
\& Williams,~B. 2010, ApJ, 713, 1322

\bibitem[{{Li \& Ye}(2000)}]{slh}
Li, W. \& Ye,~K.Q. 2000, Journal of Statistical Planning and
Inference, 90, 145 

\bibitem[{{Lin et al.}(2011)}]{lin} 
Lin,~H.; Dodelson,~S., Seo,~H.-J., Soares-Santos,~M., Annis,~J.,
Hao,~J., Johnston,~D., Kubo,~J.M., Reis,~R.R.R., \& Simet,~M. 2012,
ApJ, 761, 15 

\bibitem[{{Merloni et al.}(2012)}]{erosita} 
Merloni,~A. et al. 2012, arXiv:1209.3114 [astro-ph.CO] 

\bibitem[{{Peacock \& Dodds}(1996)}]{PD96} 
Peacock,~J.A. \& Dodds,~S.J. 1996, MNRAS, 280, L19 

\bibitem[{{Perlmutter et al.}(1999)}]{perlmutter} 
Perlmutter,~S. et al. 1999, ApJ, 517, 565

\bibitem[{{Refregier et al.}(2010)}]{euclid} 
Refregier,~A., Amara,~A., Kitching,~T.D., Rassat,~A., Scaramella,~R.,
Weller,~J. for the Euclid Imaging Consortium, 2010, arXiv:1001.0061
[astro-ph.IM] 

\bibitem[{{Riess et al.}(1998)}]{riess} 
Riess,~A.G. et al. 1998, AJ, 116, 1009
  
 \bibitem[{{Riess et al.}(2011)}]{riess_hubble} 
Riess,~A.G. et al. 2011, ApJ, 730, 119

\bibitem[{{Rudd et al.}(2008)}]{rudd} 
Rudd,~D., Zentner,~A., \& Kravtsov,~A.  2008, ApJ, 672, 19 

\bibitem[{{Santner et al.}(2003)}]{santner03} 
Santner,~T.J., Williams,~B.J., \& Notz,~W.I. 2003, The Design and
Analysis of Computer Experiments, Springer, New York

\bibitem[{{Schlegel, White, \& Eisenstein}(2009)}]{boss} 
Schlegel,~D., White,~M. \& Eisenstein,~D. arXiv:0902.4680
[astro-ph.CO] 

\bibitem[{{Schneider et al.}(2008)}]{SKHHHN} 
Schneider,~M., Knox,~L., Habib,~S., Heitmann,~K., Higdon,~D., \&
Nakhleh,~C. 2008, Phys. Rev. D, 78, 063529 
  
\bibitem[{{Schneider, Holm, \& Knox}(2011)}]{schneider}
Schneider,~M., Holm,~O., \& Knox,~L., 2011, ApJ, 728, 137 

\bibitem[{{Semboloni et al.}(2006)}]{semboloni} 
Semboloni,~E., Mellier,~Y., van Waerbeke,~L., Hoekstra,~H.,
Tereno,~I., Benabed,~K., Gwyn,~S.D.J., Fu,~L., Hudson,~M.J.;
Maoli,~R., \& Parker,~L.C. 2006, A\&A, 452, 51

\bibitem[{{Semboloni et al.}(2011)}]{semboloni11}
Semboloni,~E., Hoekstra,~H., Schaye,~J., van Daalen,~M.P., \&
McCarthy,~I.G. 2011, MNRAS 417, 2020  

\bibitem[{{Smith et al.}(2003)}]{Smith03} 
Smith,~R.E. et al.  [The Virgo Consortium Collaboration], 2003, MNRAS
341, 1311  

\bibitem[{{Springel}(2005)}]{gadget2}
Springel,~V. 2005, MNRAS, 364, 1105 

\bibitem[{{Sugai}(2012)}]{sumire} 
Sugai,~H. et al. arXiv:1210.2719 [astro-ph.IM] 

\bibitem[{{Takahashi et al.}(2012)}]{takahashi} 
Takahashi,~R., Sato,~M., Nishimichi,~T., Taruya,~A., \&
Oguri,~M. 2012, ApJ, 761, 152 
  
\bibitem[{{Taruya et al.}(2012)}]{Taruya}
Taruya, A., Bernardeau,~F., Nishimichi,~T., \& Codis~S. 2012, Phys.\
Rev.\ D 86, 103528  

\bibitem[{{van Daalen et al.}(2011)}]{vanDaalen} 
van Daalen,~M.P., Schaye,~J., Booth~C.M., \& Vecchia~C.D. 2011, MNRAS,
415, 3649  

\bibitem[{{van Engelen et al.}(2012)}]{spt} 
van Engelen,~A. et al. 2012, ApJ, 756, 142

\bibitem[{{Vikhlinin et al.}(2009)}]{vikhlinin} 
Vikhlinin,~A., Kravtsov,~A.V., Burenin,~R.A., Ebeling,~H.,
Forman,~W.R., Hornstrup,~A., Jones,~C., Murray,~S.S., Nagai,~D.,
Quintana,~H., \&  Voevodkin,~A. 2009, ApJ, 692, 1060 

\bibitem[{{Weinberg et al.}(2012)}]{weinberg} 
Weinberg,~D.H., Mortonson,~M.J., Eisenstein,~D.J., Hirata,~C.,
Riess,~A.G., \& Rozo,~E. 2012, arXiv:1201.2434 [astro-ph.CO] 

\bibitem[{{White}(2004)}]{white04} 
White,~M. 2004, Astropart. Phys., 22, 211

\bibitem[{{Wu, Zentner, \& Wechsler}(2010)}]{wu} 
Wu,~H.-Y., Zenter,~A.R., \& Wechsler,~R. 2010, ApJ, 713, 856 

\bibitem[{{Zehavi et al.}(2011)}]{zehavi}
Zehavi,~I. et al. 2011, ApJ, 726, 13

\bibitem[{{Zentner et al.}(2012)}]{zentner12}
Zentner,~A.R., Semboloni,~E., Dodelson,~S., Eifler,~T., Krause,~E., \&
Hearin,~A.P. 2012, arXiv:1212.1177 [astro-ph.CO] 

\bibitem[{{Zhan \& Knox}(2004)}]{zhanknox}
Zhan,~H. \& Knox,~L. 2004, ApJ, 616, L75   

\end{thebibliography}
\end{document}